\documentclass[twocolumn]{aastex631}

\usepackage{amsmath}

\begin{document}

\title{A Turbulent Framework for Star Formation in High-Redshift Galaxies}

\correspondingauthor{Guochao Sun} \email{guochao.sun@northwestern.edu}
\author{Guochao Sun}
\affiliation{CIERA and Department of Physics and Astronomy, Northwestern University, 1800 Sherman Ave, Evanston, IL 60201, USA}
\author{Claude-Andr\'{e} Faucher-Gigu\`{e}re}
\affiliation{CIERA and Department of Physics and Astronomy, Northwestern University, 1800 Sherman Ave, Evanston, IL 60201, USA}
\author{Jonathan Stern}
\affiliation{School of Physics and Astronomy, Tel Aviv University, Tel Aviv 69978, Israel}



\begin{abstract}

Observations of distant galaxies suggest that the physics of galaxy formation at high redshifts differs significantly from later times. In contrast to large, steady disk galaxies like the Milky Way, high-redshift galaxies are often characterized by clumpy, disturbed morphologies and bursty star formation histories. These differences between low-mass, bursty galaxies and higher-mass, steady star-forming galaxies have recently been studied in galaxy formation simulations with resolved multiphase ISM. These simulation studies indicate that while steady disk galaxies can be well-modeled as ``equilibrium disks" embedded in a distinct, hot CGM, bursty galaxies are much more dynamic and their star formation occurs in a dispersion-dominated medium that extends to halo scales, with no clear boundary between the ISM and the CGM. We develop an analytic framework to model star formation in bursty galaxies that are not adequately modeled as equilibrium disks. The framework approximates the gas in low-mass halos as a continuous, supersonically turbulent medium with large density fluctuations. Star formation occurs locally in the high-density tail of a roughly lognormal density distribution. This is analogous to turbulent models of star formation in molecular clouds, but here applied on inner CGM scales. By comparing with galaxy formation simulations from the FIRE project, we show that this framework can be used to understand star formation efficiencies and radial profiles in halos. The turbulent framework shows explicitly how the instantaneous galaxy-averaged star formation efficiency can be relatively low even if the local efficiency in dense gas approaches unity. 

\end{abstract}

\keywords{Galaxy formation (595) --- High-redshift galaxies (734) --- Hydrodynamical simulations (767) --- Interstellar medium (847) --- Star formation (1569) --- Theoretical models(2107)}


\section{Introduction}

Star formation is a critical yet complex process that creates stellar populations as the building blocks of galaxies, provides various kinds of radiative, kinetic, and chemical feedback that regulate galaxy evolution, and sources the metagalactic radiation background imposing large-scale effects on galaxies and the intergalactic medium (IGM). The intricate nature of star formation in galaxies can be attributed to the immense dynamical range and diverse physical processes involved, including the interplay between gravity and turbulence, magnetic fields, stellar feedback, chemical enrichment, and environmental effects, among others \citep{McKeeOstriker2007,Ostriker2010,AgertzKravtsov2015,HaywardHopkins2017,Semenov2018}. Consequently, understanding the physics that governs star formation in galaxies lays the foundation for a robust theory of the formation and evolution of galaxies and the IGM across cosmic times \citep{MD2014,Somerville2015,McQuinn2016}. 

The relationship between galaxy-scale star formation and the physical properties of star-forming galaxies (SFGs) has been studied in substantial detail both observationally and theoretically in past decades. Deep, spatially resolved observations by telescopes such as the Hubble Space Telescope (HST) at rest-frame UV wavelengths tracing recent star formation have revealed an increased fraction of SFGs with clumpy/disturbed morphology at lower mass and/or higher redshift \citep{Guo2015,Huertas-Company2016,Sok2022,Sattari2023}, whereas measurements of star formation rate (SFR) indicators probing SFR variations on different timescales (e.g., H$\alpha$ vs.\ UV continuum luminosities) have provided evidence that lower mass/higher redshift galaxies have star formation histories (SFHs) that are more strongly time-variable, or bursty \citep{Weisz2012,Emami2019,Faisst2019}. These observations have led to speculation about whether transitions in the SFR variability (bursty vs.\ steady) and morphology (clumpy/disturbed vs.\ rotationally dominated) of SFGs are physically connected. 

Cosmological hydrodynamical simulations with detailed star formation and feedback physics implemented offer invaluable insights into possible physical factors that distinguish bursty, morphologically disturbed SFGs from massive ones similar to local spiral galaxies characterized by stable, rotationally supported disks and time-steady SFHs. Interestingly, several recent analyses of Milky-Way analogs/progenitors in high-resolution galaxy formation simulations with a resolved multiphase interstellar medium (ISM) and multichannel stellar feedback have identified potential connections between disk formation and the bursty-to-steady transition of galaxy-scale star formation \citep[e.g.,][]{Stern2021,Yu2021,Yu2023,Gurvich2023,Semenov2024Early,Semenov2024MWD}. The exact physical drivers of such a connection and the specific physical conditions under which the morphological and star formation mode transitions take place are still debated and possibly interrelated. Some factors identified include the depth and concentration of the gravitational potential, the confinement of feedback-driven outflows by gravity and/or the hot circumgalactic medium (CGM), and how angular momentum is delivered to forming galaxies by cold vs.\ hot mode accretion \citep{Martizzi2020,Stern2021,Dekel2020,Hafen2022,Byrne2023,Hopkins2023,Semenov2024FGD2}. Although we do not yet have a full causal theory, the theoretically predicted distinctions in the early- and late-phase evolution SFGs appear to explain several properties of observed galaxies. These theoretical developments should therefore be taken into account when modeling low-mass SFGs, which become increasingly abundant at high redshift.

Since its launch, the James Webb Space Telescope (JWST) has observed the high-redshift ($z\gtrsim6$) universe at unprecedented depth and resolving power, allowing the properties of early SFGs to be scrutinized at epochs when the universe was characterized by significantly younger age, higher average density, and more rapid structure formation. Among the most intriguing discoveries JWST has made so far are the large number of galaxies with numerous star-forming clumps and irregular morphologies \citep[e.g.,][]{Kartaltepe2023,Fujimoto2024,Hainline2024} and the prevalence of galaxies with highly bursty SFHs \citep[e.g.,][]{Ciesla2024,Dressler2024,Endsley2024} in the early universe. Observations have also reported a high number density of UV-bright galaxies at $z>10$ and shallow evolution of the cosmic UV luminosity density \citep{Bouwens2023,Donnan2023,Harikane2024}, which have motivated re-examinations of the key aspects of star formation in galaxies, such as the possibility of an increased star formation efficiency (SFE) due to inefficient feedback \citep{Dekel2023,Li2024} and implications of highly bursty SFHs \citep{Mason2023,MirochaFurlanetto2023,Shen2023,Sun2023_seen,Sun2023,Kravtsov2024}. 

It is noteworthy that these high-redshift SFGs are typically well within the mass regime ($M_\mathrm{\star} \lesssim 10^{10}\,M_{\odot}$ or $M_\mathrm{vir} \lesssim 10^{12}\,M_{\odot}$) where bursty star formation and disturbed/irregular morphologies are predicted in simulations. Morphological and kinematic analyses of the gas content of bursty SFGs in these simulations have shown that the support of the cold interstellar gas against gravity in these galaxies is primarily supplied by quasi-isotropic dispersion and turbulence-like bulk flows (powered inflows and outflows) instead of rotation \citep{Stern2021,Gurvich2023,McCluskey2024}. Moreover, galaxies in this mass regime are  predicted to be surrounded by a ``pre-virialized'' CGM, with order-unity mass fractions in cool gas \citep[]{Stern2021,Stern2021DLA,Gurvich2023} especially at inner halo radii. \cite{Kakoly2025} showed that such cool CGM are  characterized by dispersive, turbulent-like velocity fields and wide lognormal density distributions, in contrast with the bi-modal density distributions expected in the standard ``multi-phase" CGM picture when cooling times are long.\footnote{In galaxy formation simulations with realistic feedback, the outer CGM can be heated early on by blast waves and so can become hot even in relatively low-mass halos, even though the inner CGM remains cold and turbulent due to the short cooling times. This gives rise to the notion of ``outside-in'' virialization and the time of ``inner CGM virialization'' (ICV), when CGM virialization completes and the boundary conditions of the central galaxy change relatively abruptly \citep[][]{Stern2020, Stern2021}.} Indeed, in this regime, simulations predict there is no clear boundary between the ISM and the CGM. This suggests a turbulent early stage of galaxy formation, preceding the emergence of steady disks, a picture for which there is also empirical evidence from Galactic archaeology \citep[see e.g.,][]{BelokurovKravtsov2022,Conroy2022,XiangRix2022}. 

The physical state of the ISM/CGM system predicted by the simulations referenced above raises important questions about how to model galaxies in analytic and semi-analytic frameworks. 
A commonly adopted framework involves ``equilibrium disk'' models, in which star formation is self-regulated in quasi-steady-state galactic disks. 
In these models, feedback from star formation pressurizes the ISM and the SFR adjusts such that the ISM pressure supports the weight of the overlying gas \citep[e.g.,][]{Thompson2005,Ostriker2010,OstrikerShetty2011,CAFG2013}. ``Equilibrium disk" models have proved highly successful at explaining various properties of disk galaxies, including the Kennicutt-Schmidt relation between the SFR and gas surface densities \citep[e.g.,][]{Kennicutt1998, Genzel2010}. 
Moreover, the simplified assumptions made in analytic formulations have been validated in a variety of more self-consistent simulations of disk galaxies \citep[e.g.,][]{ShettyOstriker2012, Kim2013, Benincasa2016, Gurvich2020, Ostriker2022}. 
For these reasons, variants of such disk models are often used to model galaxies, including at high redshift \citep[e.g.,][]{Dekel2013, CAFG2018, Furlanetto2021}. However, the observations and simulations we have reviewed call into question the validity of this approach in the regime of low-mass, bursty galaxies, which become increasingly dominant at high redshift. \cite{Gurvich2023}, in particular, explicitly discussed the limitations of equilibrium disk models for modeling the bursty progenitors of Milky Way-like galaxies.

In this paper, we develop a new approach to model bursty SFGs, such as those that likely prevail in the high-$z$ universe. 
Motivated by the cool, dispersion-dominated ISM-CGM found in high-resolution cosmological simulations for such galaxies, we introduce an analytic framework that approximates the gas in and around bursty SFGs as described by a turbulent flow. 
In realistic settings, the turbulence has a broad range of temperatures, but the large cool gas filling fractions imply that large density fluctuations are created by the turbulence, similar to more idealized supersonic turbulence. 
The framework we propose builds on many previous models for the formation of individual stars in molecular clouds \citep[][]{PadoanNordlund2002,KrumholzMcKee2005,McKeeOstriker2007,FederrathKlessen2012}, but here extended to the scale of the galaxy host halo. 
In this framework, the turbulent gas density field is approximated by a roughly lognormal probability density function (PDF) whose width is set by the velocity dispersion of the gravitational potential of the host halo and star formation occurs in the high-density tail of the PDF. By validating this framework against cosmological zoom-in simulations from Feedback in Realistic Environments (FIRE) project,\footnote{\url{https://fire.northwestern.edu}} we demonstrate that various star formation properties predicted in detailed simulations, such as the total SFR, SFR radial profiles, galaxy sizes, and the star formation efficiencies, are reasonably well reproduced in our framework despite many simplifications. 
We argue that this framework provides useful insights into the physical nature of bursty galaxies and lays out a new path for analytically modeling galaxy formation that is more realistic than equilibrium disk models in the bursty regime. 
This is especially relevant at high redshift, which we focus on in this paper, but the framework likely generalizes to low-mass galaxies that remain bursty to later times.   

The remainder of this paper is organized as follows. In Section~\ref{sec:model-turbulence}, we present the analytic framework for modeling star formation in gaseous halos with turbulence-driven density fluctuations. In Section~\ref{sec:validation}, we demonstrate the general validity of the turbulent framework and assess its limitations by comparing the gas and star formation properties it predicts against measurements of galaxies in the \textit{High-Redshift} suite of the FIRE-2 simulations. In Section~\ref{sec:implications}, we contextualize the significance of our turbulent framework and in particular use it to discuss its implications for the galaxy-integrated SFE. We discuss limitations and useful extensions of the presented framework, before concluding in Section~\ref{sec:conclusions}. Throughout, we adopt a flat $\Lambda$CDM cosmology consistent with \citet{Planck_2016}.

\section{Modeling star formation from a turbulent density field} \label{sec:model-turbulence}

\begin{figure*}
 \centering
 \includegraphics[width=\textwidth]{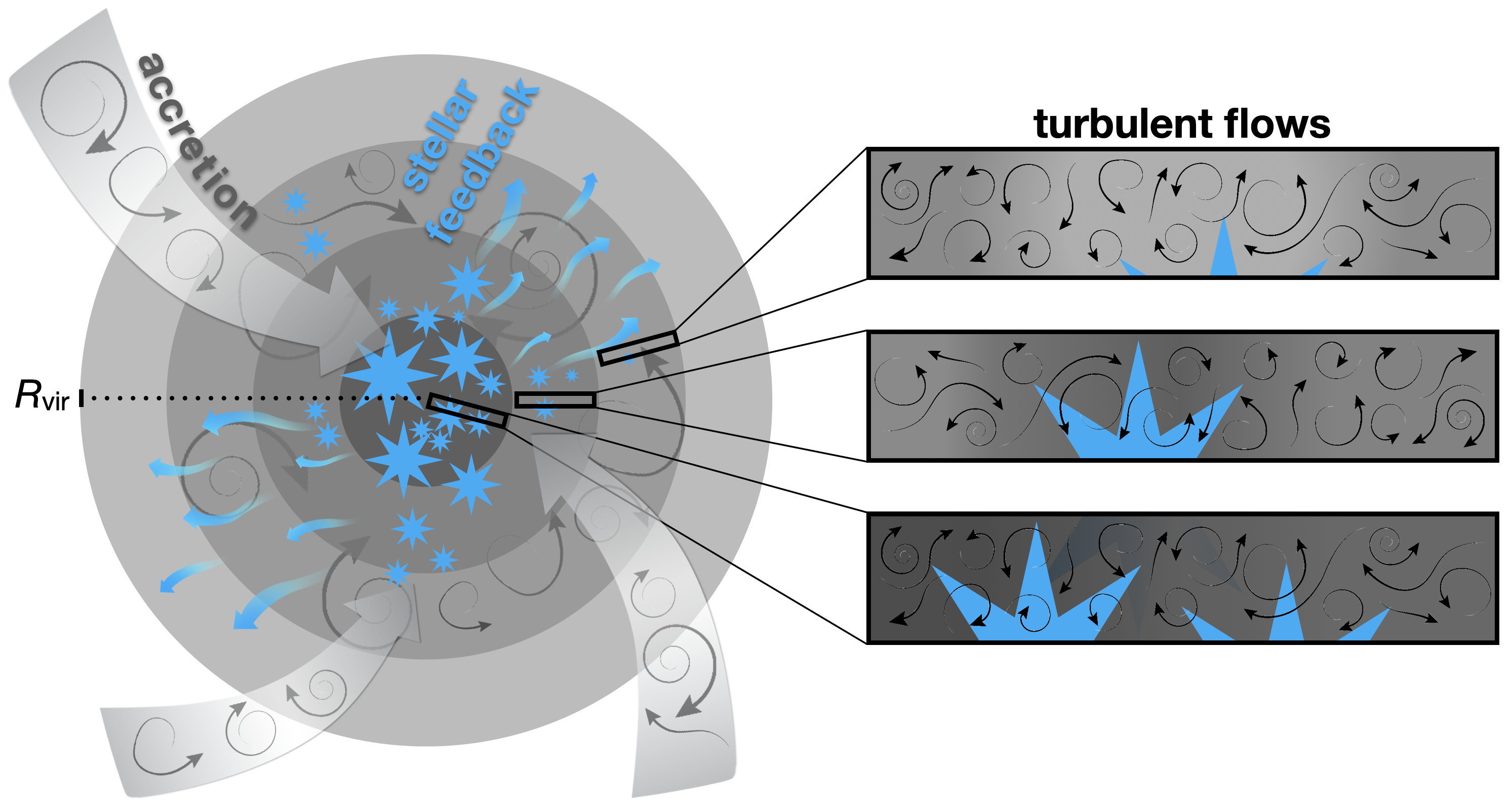}
 \caption{A schematic diagram of the turbulent framework presented in this work. In the early, bursty stage of their formation, low-mass/high-$z$ SFGs do not form stable, rotationally supported gas disks surrounded by a quasi-static hot CGM. Instead, stellar feedback and gas accretion drive strong supersonic turbulence extending out to the host halo, which does not thermalize due to the short cooling times especially the inner half halo where the majority of star formation takes place. In this regime, there is no clear distinction between ISM and CGM flows. Star formation across the halo can thus be approximated by combining the mean gas density profile and the gas density contrast distribution set by the turbulence properties with criteria for star formation in the high-density tail of the distribution.}
 \label{fig:sketch}
\end{figure*}

In this section, we present a simple analytic model for describing star formation in low-mass/high-$z$ bursty galaxies by specifying the density profile (Section~\ref{sec:model-turbulence:profile}) and density contrast PDF (Section~\ref{sec:model-turbulence:distribution}) of their gas reservoir, together with criteria for star formation that mainly involve a density threshold (Section~\ref{sec:model-turbulence:sf}). As illustrated schematically in Figure~\ref{fig:sketch}, this analytic framework describes the gaseous halo of SFGs in the bursty stage as a supersonically turbulent density field, where a clear distinction between the ISM and the CGM is absent and the gas pressure support throughout most of the halo is dominated by turbulence driven by mechanisms such as stellar feedback and gas inflows \citep{KrumholzBurkhart2016}. 

\subsection{Density Profile of Galaxy Host Halos} \label{sec:model-turbulence:profile}

Considering spherically symmetric dark matter halos whose internal mass distribution can be fully described by a density profile $\rho(r)$ for a given radial distance $r$, we can take the simplest plausible assumption that the gas distribution traces that of dark matter by a constant factor $f_\mathrm{gas}$, namely $\rho_\mathrm{gas}(r)=f_\mathrm{gas} \rho(r)$. To a decent approximation for the purpose of this work, the density profile of the host halo of virial radius $R_\mathrm{vir}$ in dynamical equilibrium follows that of a truncated and softened isothermal sphere
\begin{equation}
\rho_\mathrm{gas}(r) = f_\mathrm{gas} \rho(r) = \frac{f_\mathrm{gas} \sigma_{v}^2}{2 \pi G (r+r_\mathrm{s})^2},\hspace{0.5cm} r \leq R_\mathrm{vir}
\label{eq:tsip}
\end{equation}
where $r_\mathrm{s} \ll R_\mathrm{vir}$ is a small softening parameter acting at small halo radii to account for the impact of e.g., stellar feedback on the gas distribution over time. Equation~(\ref{eq:tsip}) then implies
\begin{equation}
\frac{\mathrm d M}{\mathrm d r} = 4\pi r^2 \rho(r) \simeq \frac{2 \sigma_{v}^2}{G},
\label{eq:mgrad}
\end{equation}
where $\sigma_{v}$ is the 1D dark matter velocity dispersion and the truncation is defined at the virial radius, $R_\mathrm{vir}$, which is taken to be the radius where the spherical overdensity of a halo of mass $M_\mathrm{vir}$ is $\Delta_\mathrm{vir}$ times the mean matter density $\rho_\mathrm{m}(z) = \Omega_\mathrm{m}(z)\rho_\mathrm{crit}$. The overdensity of a virialized halo $\Delta_\mathrm{vir} = 18\pi^2 + 82x - 39x^2$, with $x=\Omega_\mathrm{m}(z)-1$ following \citet{BryanNorman1998}. Under these assumptions, the virial velocity of the halo is
\begin{equation}
v_\mathrm{vir}=\sqrt{GM_\mathrm{vir}/R_\mathrm{vir}} = \sqrt{2} \sigma_{v}. 
\end{equation}
Such $\rho\propto r^{-2}$ profiles are indeed seen in FIRE simulations of high redshift galaxies with pre-virialized CGM \citep{Stern2021DLA}, and can also be derived directly from the fluid equations assuming turbulence provides the main support against gravity instead of thermal pressure \citep{Goldner2025}.

\subsection{Density Distribution of Turbulent Gas} \label{sec:model-turbulence:distribution}

Cosmological hydrodynamical simulations with resolved multiphase ISM and stellar feedback predict the interstellar gas of SFGs in the early, bursty stage of formation to be highly dynamic and turbulent (e.g., \citealt{Gurvich2023}; \citealt{Semenov2024Early}; see Section~\ref{sec:validation}), with the kinetic pressure from turbulence and bulk flows dominating the gas energetics up to a significant fraction of the halo scale. Thus, here we consider the analytic picture of star formation in supersonically turbulent molecular clouds that involves a log-normal-like gas density contrast distribution and attempt to generalize it to halo scales to describe the simulated bursty and clumpy SFGs in the early universe. The density distribution of isothermal supersonic turbulence has been shown to be well approximated by a log-normal PDF \cite[see e.g.,][]{PVS1998,OSG2001,Kritsuk2007,PN2011}, as expected from density fluctuations built up from a hierarchical process that is randomly and identically additive in logarithmic density. Specifically, the (mass-weighted) density contrast PDF for the turbulent gas is given by
\begin{equation}
P_{M}(y) \mathrm d y = \frac{1}{\sqrt{2\pi \mathcal{S}}} \exp \left[ -\frac{(y - y_0)^2}{2\mathcal{S}} \right] \mathrm d y, 
\label{eq:pdf_ln}
\end{equation}
where $y \equiv \ln(\rho_\mathrm{gas}/\langle \rho_\mathrm{gas} \rangle)$ is the logarithmic density contrast with respect to the volume-weighted average density $\langle \rho_\mathrm{gas} \rangle$ and $y_0 = \mathcal{S}/2$ as required by $\langle y \rangle = 0$. 

The variance of the density contrast distribution, $\mathcal{S}$, is found to depend on physical properties of the turbulence, including (1) the Mach number, $\mathcal{M}$, (2) the relative importance of the two mechanisms of kinetic energy injection that drives the turbulence (solenoidal or ``divergence-free" vs. compressive or ``curl-free") captured by a parameter $0<b<1$, and (3) whether or not magnetic fields are present \citep{Molina2012,FK2013}. For purely hydrodynamical isothermal 3-dimensional turbulence (e.g., without the influence of magnetic fields), 
\begin{equation}
\mathcal{S} \approx \ln (1 + b^2 \mathcal{M}^2), 
\label{eq:S}
\end{equation}
with $b \sim 1/3, 1$, and $2/5$ for pure solenoidal driving, pure compressive driving, and a random, natural mixture of both modes \citep{Federrath2010}, respectively. 

\cite{Kakoly2025} showed that lognormal density distributions appear also at halo scales when cooling times of the hot gas are short, so cool gas dominates the mass, the mean sound speed is low and thus CGM turbulence is supersonic. We can analytically estimate the variance by assuming a turbulent Mach number 
\begin{equation}
\mathcal{M} \approx \frac{\sqrt{3} \sigma_{v,\mathrm{gas}}^\mathrm{1D}}{c_\mathrm{s}} \approx \frac{\sqrt{3} \sigma_{v}}{c_\mathrm{s}},
\label{eq:mach}
\end{equation}
where the speed of sound $c_\mathrm{s} = \sqrt{\gamma k_\mathrm{B} T / \mu m_\mathrm{p}} \sim$ a few to 10 $\,\mathrm{km/s}$ in typical warm neutral/ionized medium which dominates the mass budget when the cooling time of hot gas is shorter than the local free-fall time and $\gamma = 5/3$ for monoatomic gas. Because our analytic model does not predict $b$ from first principles, we calibrate the choice of $b$ against measurements of the FIRE simulations (see Section~\ref{sec:validation}), from which we find that $b = 0.8$ provides good approximations to the gas density contrast PDFs. This is consistent with the value found by \citet{Kakoly2025} for Milky-Way-like galaxies in the early supersonic stage from the FIRE simulations. 

In self-gravitating gas, the high-density tail of the density PDF is expected to deviate from a lognormal and approach a power law upon gravitational collapse \citep[see e.g.,][and references therein]{BurkhartMocz2019}. We examine this possibility in Appendix~\ref{sec:appendix-ln+pl} for the high-redshift FIRE simulations in this paper by fitting the simulated PDF with a piecewise model for the power-law transition. No clear evidence is found for a strong power-law modification of the high-density tail in the inner halo, where most of the star formation occurs, for the fiducial runs with local star formation efficiency $\epsilon_{\rm SF}=1$. In Appendix \ref{sec:appendix-altmodels}, we show that significant deviations from a lognormal PDF emerges when $\epsilon_{\rm SF}$ is lowered to 0.1 or 0.01, allowing gas to collapse longer before turning into stars.

\subsection{Star Formation in Turbulent Gas} \label{sec:model-turbulence:sf}

\begin{figure}
 \centering
 \includegraphics[width=\columnwidth]{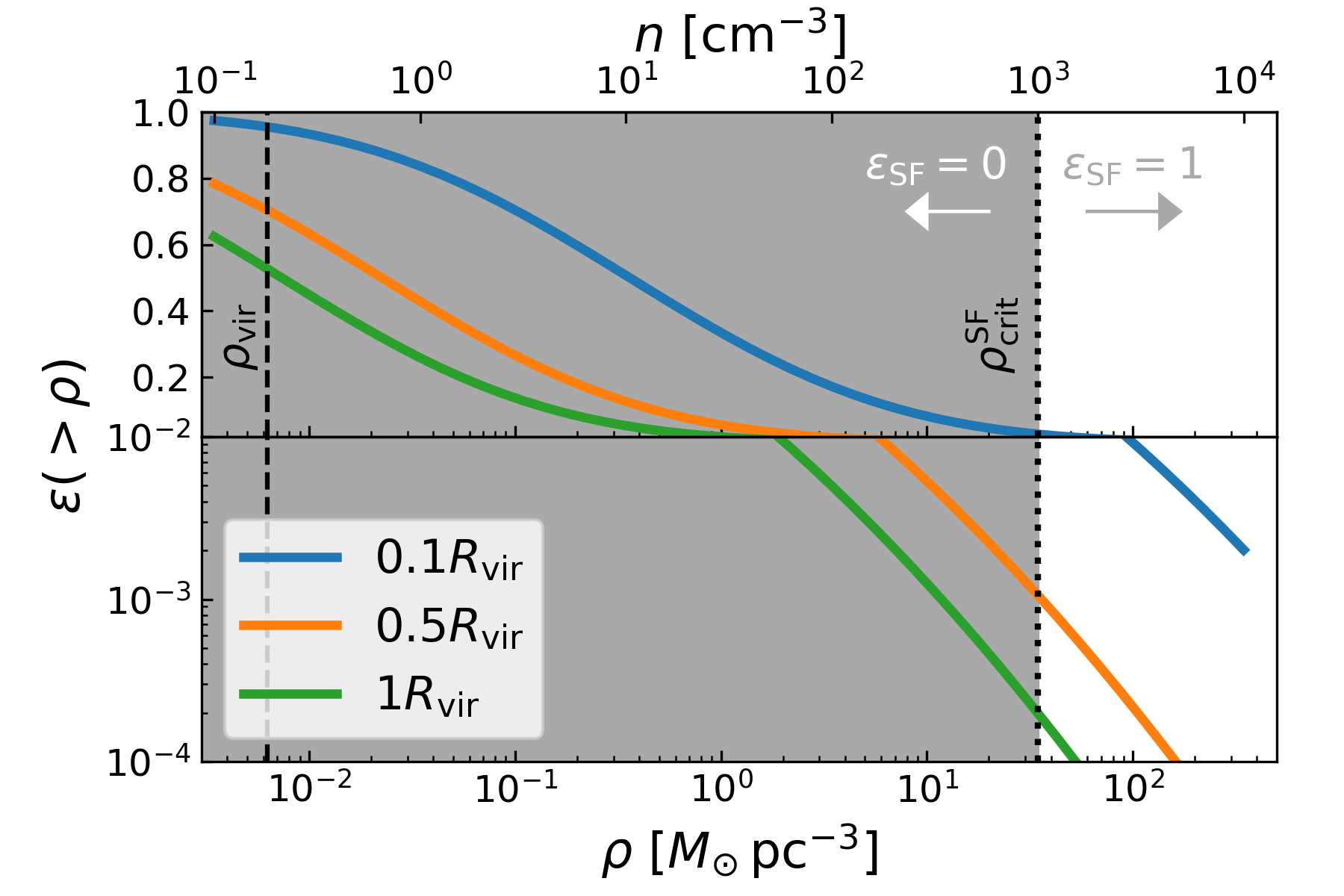}
 \caption{The mass fraction of gas above a given density threshold $\rho$ at different halo radii predicted by our analytic model for log-normal density contrast distribution. Marked by the vertical lines are the virial density $\rho_\mathrm{vir}=\Delta_\mathrm{vir}(z)\rho_\mathrm{c}(z)$ at $z=8$ and the critical density $\epsilon_\mathrm{SF}$ above which the SFE per free-fall time jumps from 0 to 1 (corresponding to $n_\mathrm{crit}=10^3\,\mathrm{cm^{-3}}$). Only a small fraction of the total gas mass can form stars and the fraction decreases with the radial distance.}
 \label{fig:efficiency_vs_density}
\end{figure}

While the exact conditions for star formation depend on various physical processes, galaxy formation models often adopt a critical gas number density, $n_\mathrm{crit}$, as a simple threshold criterion, which is also a commonly used subgrid prescription in numerical simulations. By integrating the (mass-weighted) density PDF over the regime of star-forming gas, we can evaluate the mass fraction of turbulent gas that forms stars
\begin{equation}
f^\mathrm{SF}_\mathrm{gas} = \int_{y_\mathrm{crit}=\ln(\rho_\mathrm{crit}/\langle \rho \rangle)} P_{M}(y) \mathrm d y, 
\label{eq:f_sf}
\end{equation}
where $\rho_\mathrm{crit}=\mu m_\mathrm{p} n_\mathrm{crit}$ for the mean molecular mass $\mu$ and the proton mass $m_\mathrm{p}$. Combining this star-forming gas mass fraction with the gas mass gradient and the star formation efficiency $\epsilon_\mathrm{SF}$ per free-fall time $t_\mathrm{ff}$ evaluated at $n_\mathrm{crit,gas}$, we can express the SFR profile of the turbulent galaxy host halo as
\begin{equation}
\frac{\mathrm d \dot{M}_{\star}}{\mathrm d r} = \frac{\mathrm d M_\mathrm{gas}}{\mathrm d r} f^\mathrm{SF}_\mathrm{gas} \frac{\epsilon_\mathrm{SF}}{t_\mathrm{ff}},
\label{eq:sfrgrad}
\end{equation}
where the free-fall timescale equals to
\begin{equation}
t_\mathrm{ff} = \sqrt{\frac{3\pi}{32G \langle \rho \rangle}} \approx \sqrt{\frac{3\pi}{32 G \mu m_\mathrm{p} n_\mathrm{crit}}}. 
\end{equation}
The mass fraction predictions from our analytic arguments are illustrated in Figure~\ref{fig:efficiency_vs_density} as a function of density threshold or halo radius. Note that as a result of the gas density profile and the critical density for star formation, only a small fraction of the total gas mass actually forms stars and the fraction drops rapidly with increasing radial distance. In future work, it would be interesting to analyze how the present framework could be modified to use more detailed and physically grounded star formation criteria, such as ones in which star formation occurs only in self-gravitating gas through a density threshold that depends on the virial parameter $\alpha_\mathrm{vir}$ \citep{KrumholzMcKee2005}. 

\begin{figure*}
 \centering
 \includegraphics[width=\textwidth]{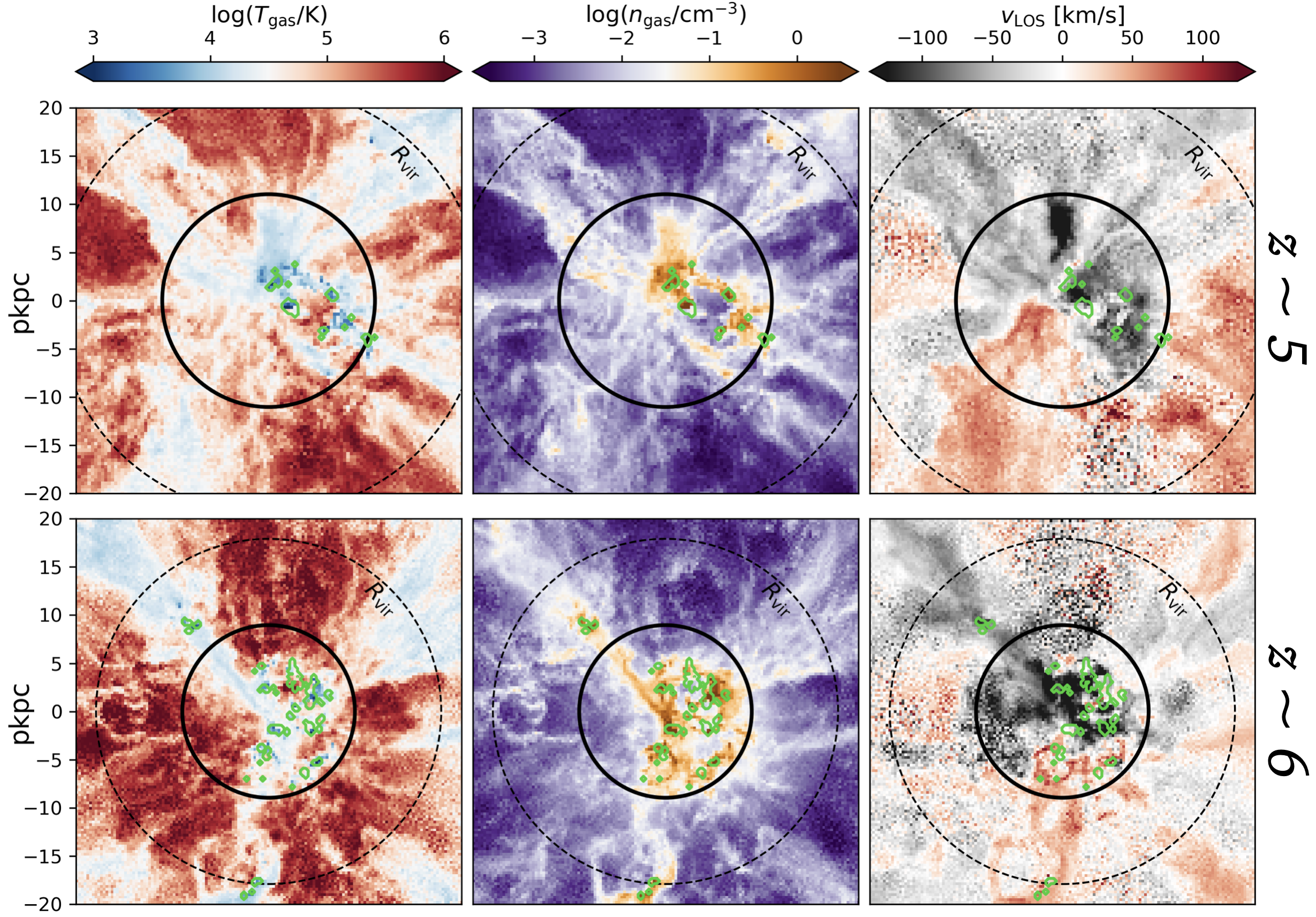}
 \caption{Gas temperature, number density, and line-of-sight velocity projections of the simulation $\texttt{z5m11h}$ in two snapshots at $z\sim5$ and 6, respectively. For temperature and density, the mass-averaged logarithmic values are shown along each projection, whereas for velocity the linear mass-averaged values are shown. The temperature and density distributions are characterized by large spatial fluctuations. A mixture of cold ($T_\mathrm{gas}<10^3\,$K), warm ($10^3<T_\mathrm{gas}<10^5\,$K), and hot ($T_\mathrm{gas}>10^5\,$K) gas extends to large halo radii, with no clear boundary between the ISM and CGM. The gas motions are characterized by a combination of dispersion and bulk flows, with dynamically subdominant rotation and thermal pressure near the center (see also Fig. \ref{fig:comp_energetics}). For reference, contours for the mass surface density of young stars (formed in the past 10\,Myr) $\Sigma_{\star, \rm young}=10^4\,M_{\odot}\,\mathrm{kpc}^{-2}$ are overlaid to show the correspondence between cold, dense gas and recent star formation. The black dashed (solid) circles indicate the virial radius (half virial radius) of the galaxy host halos. Most of the star formation occurs in the inner half of the halo (as quantified in Figure~\ref{fig:subhalo_frac}) owing to the higher average gas densities, but there is no well-defined boundary and some star formation takes place outside this radius.
}
 \label{fig:maps}
\end{figure*}

From Equation~(\ref{eq:sfrgrad}), it is also possible to estimate the stellar mass of the galaxy by
\begin{equation}
M_{\star}(<r) \simeq \int_{0}^{r} (1-R) \frac{d \dot{M}_\mathrm{\star}}{d r'} \frac{t_{\star}}{H(z)} d r', 
\label{eq:mstar}
\end{equation}
where $R \approx 0.25$ is the return fraction due to stellar mass loss,  $H(z)$ is the Hubble parameter, and $t_{\star}$ (dimensionless) defines a characteristic timescale, $t_{\star}/H(z)$, of star formation. Previous studies have found that $t_{\star} \sim 0.3$ provides a good match to the observed galaxy UV luminosity function at $z \gtrsim 6$ \citep{Park2019,MasRibas2023}. An important clarification here is that while Equation~(\ref{eq:mstar}) holds as a rough estimate of the stellar mass formed in a halo, it implicitly assumes a fixed known $f_\mathrm{gas}$. Even though the baryonic content of halos, especially the low-mass ones, can be strongly depleted by feedback \citep[e.g.,][]{Hafen2019,Sorini2022}, the physical processes that set the baryon fractions in halos (such as the full time history of inflows and outflows) are external to our model. In our framework, we assume a fixed fraction $f_\mathrm{gas} = 0.1$, which well reproduces the simulated gas density profiles shown in Figure~\ref{fig:gas_profiles} (see also Section~\ref{sec:conclusions} for more discussion of this assumption), and study how turbulence determines the fraction of gas that forms stars as a function of halo radius. This allows us to predict the SFR and related properties such as its spatial distribution in the halo, but not to predict the galaxy-scale time-integrated SFE in halos such as characterized by the  stellar mass-halo mass relation (see Section~\ref{sec:implications} for further discussion).

\section{Validating the Turbulent Framework in Simulations} \label{sec:validation}

\subsection{Simulations}

In order to validate the turbulent framework described in Section~\ref{sec:model-turbulence}, we compare the model predictions to the relevant gas and star formation properties of high-$z$ galaxies predicted by cosmological zoom-in simulations. The simulations we analyze here are taken from a collection of reruns of the high-redshift suite FIRE-2 simulations first introduced by \citet{Ma2018Morph,Ma2018LF,Ma2019}. Using initial conditions identical to the original runs, these simulations were created using the GIZMO code \citep{Hopkins2015} with the hydrodynamic equations solved by GIZMO's meshless-finite mass (MFM) hydro solver, after fixing a cosmic ray heating issue affecting CGM/IGM gas thermodynamics in high-$z$ halos in the original runs. While this cosmic ray heating issue has little impact on the star formation properties of the galaxies that have been studied in previous work, it significantly alters the gas temperature distribution on CGM/IGM scales, which is more important for the present work. We provide a brief overview of the key aspects of the simulations used for validation below, while noting that detailed information about the FIRE-2 physical models can be found in \citet{Hopkins2018}. 

Radiative cooling of gas is traced in the simulations over $10$ to $10^{10}\,$K, including low-temperature fine-structure and molecular cooling, as well as high-temperature metal-line cooling separately tracked by species. The gas is irradiated by a redshift-dependent but homogeneous ionizing background \citep{CAFG2020}\footnote{Note that this has been updated from the UV background model \citep{CAFG2009} used in original FIRE-2 simulations \citep{Hopkins2018,Ma2018Morph,Ma2018LF}.}. The star formation criteria, taken from \citet{Hopkins2013}, convert the gas particle into a star particle with 100\% efficiency per local free-fall time when it is sufficiently dense ($n_\mathrm{gas} \geq 1000\,\mathrm{cm}^{-3}$), self-gravitating and self-shielding. Following star formation, multiple channels of stellar feedback are implemented, which regulates the galaxy-scale time-integrated SFE to $M_{\star}/f_\mathrm{b} M_\mathrm{vir} \sim1$--$10\%$ (or less) depending on the halo mass. Mechanisms of stellar feedback implemented include: (1) energy, momentum, mass, and metal injections from supernovae (Type II and Ia) and stellar winds, (2) photoionization and photoelectric heating, and (3) radiation pressure with both local and long-range components.  

\begin{figure*}
 \centering
 \includegraphics[width=\textwidth]{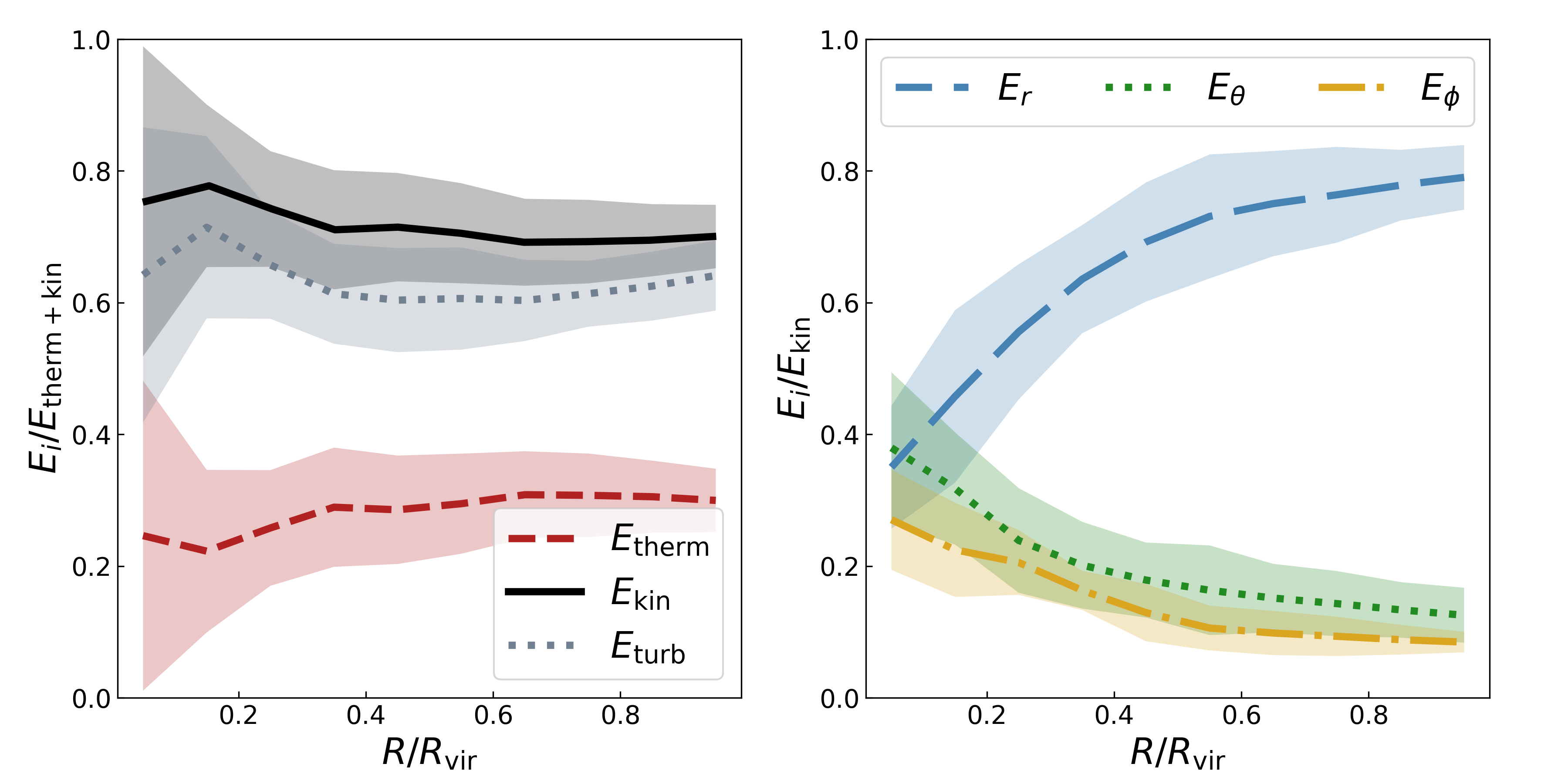}
 \caption{Comparison of the thermal energy ($E_\mathrm{therm}$), the total kinetic energy ($E_\mathrm{kin}$) and its three spherical components ($E_{R}$, $E_{z}$, and $E_{\phi}$), and the turbulent energy based on velocity dispersions ($E_\mathrm{turb}$, defined as in Equation~(\ref{eq:Eturb})) for the halo gas in the simulation $\texttt{z5m11h}$ as a function of halo radius. The average and $1\sigma$ dispersion measured from snapshots over $5<z<6$ are represented by the curves and shaded bands, respectively. Left: gas kinetic energy (the majority of which comes from turbulent motions) dominates over thermal energy across the halo. Right: the gas shows no sign of significant rotational support, i.e., the polar and azimuthal components, $E_{\theta}$ and $E_{\phi}$, never dominates. The three components of kinetic energy are comparable (quasi-isotropic) at small radii, whereas $E_R$ becomes increasingly significant at larger radii, likely as a consequence of turbulence driven by inflows/outflows.}
 \label{fig:comp_energetics}
\end{figure*}

We focus our subsequent analyses on the simulation of a typical low-mass galaxy, $\texttt{z5m11h}$, which reaches a halo (stellar) mass of $8.2 \times 10^{10} \,M_{\odot}$ ($6.7 \times 10^8\, M_{\odot}$) at $z=5$, but note that 11 additional simulations reruns ($\texttt{z5m11d}$, $\texttt{z5m11e}$, $\texttt{z5m11f}$, $\texttt{z5m11g}$, $\texttt{z5m12a}$, $\texttt{z5m12c}$, $\texttt{z5m12d}$, $\texttt{z5m12e}$, $\texttt{z7m12a}$, $\texttt{z7m12b}$, $\texttt{z7m12c}$) with the same updates are also created and used at various steps of the validation to enhance statistical robustness (see e.g., Figures~\ref{fig:S_comp} and \ref{fig:sfe}). The dark matter and baryonic mass resolutions of these simulations are $3.9\times10^4\,M_{\odot}$ and $7.1\times10^3\,M_{\odot}$, respectively. The gravitational softening length in physical units is fixed to 42\,pc (2.1\,pc) for the dark matter (stars) and is adaptive for gas with a minimum of 0.42\,pc in densest regions. Snapshots of the simulations are saved every $\sim15\,$Myr. For each simulation snapshot, we examine only the primary galaxy/halo located at the center of the zoom-in region. The primary halo is identified by applying a modified version of the \textsc{rockstar} halo finder \citep{Behroozi2013} called \textsc{rockstar-galaxies}\footnote{\url{bitbucket.org/awetzel/rockstar-galaxies}} to the dark matter particles, after which gas and star particles in different radial bins measured from the identified halo center are selected for subsequent analysis. 

\begin{figure}
 \centering
 \includegraphics[width=\columnwidth]{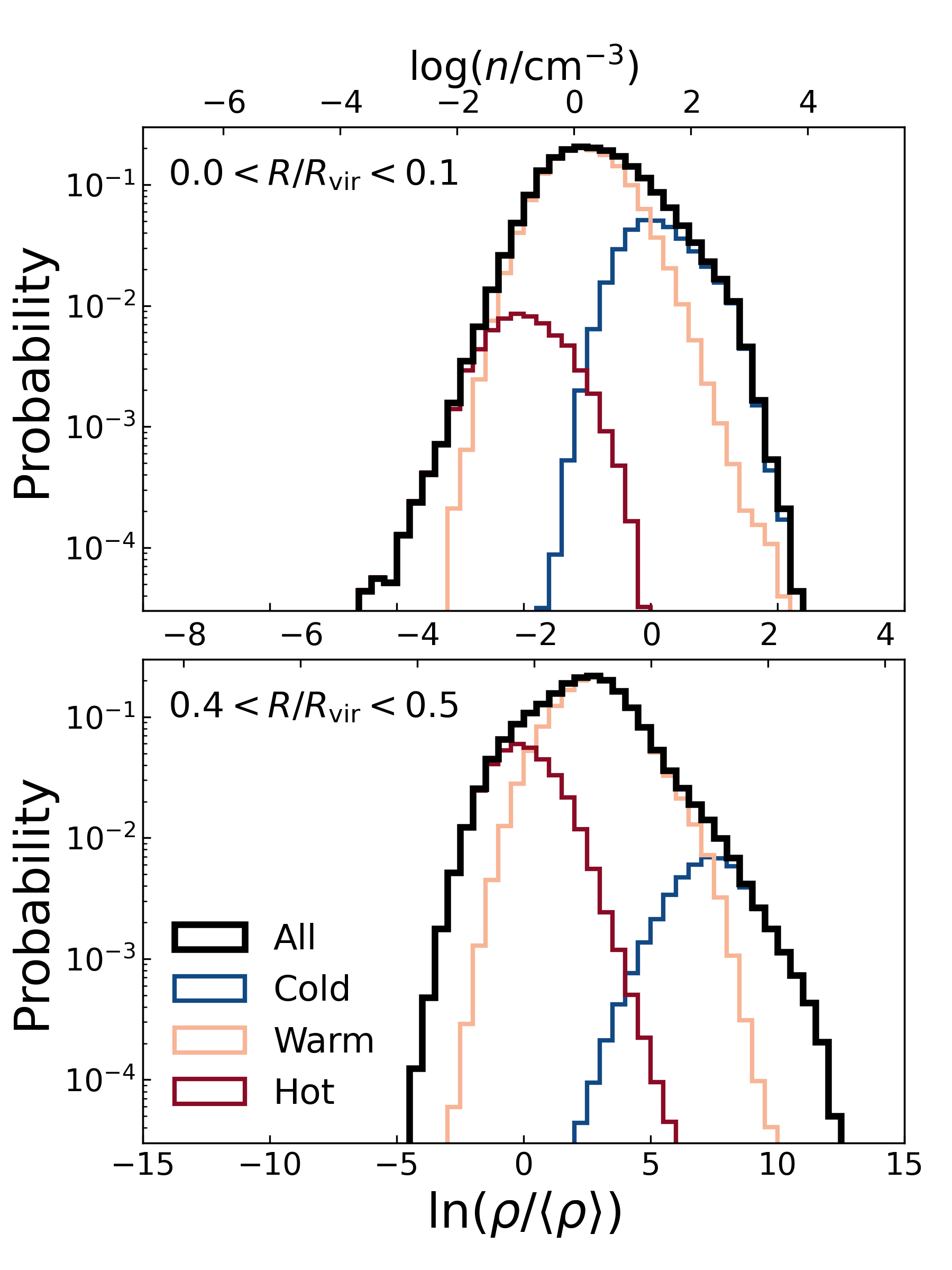}
 \caption{A comparison of the mass-weighted PDFs of the density contrast $y=\ln{(\rho/\langle \rho \rangle)}$ for gas in different temperature ranges (cold: $T_\mathrm{gas}<10^3\,$K; warm: $10^3<T_\mathrm{gas}<10^5\,$K; hot: $T_\mathrm{gas}>10^5\,$K) at two different halo radii. The PDFs are evaluated using the combination of gas particle data in the corresponding radial bin taken from the 16 snapshots of the simulation \texttt{z5m11h} over $5<z<6$, normalized according to the mass fraction of each temperature range with respect to the total gas mass. Although the PDFs vary across different gas temperatures, the total $P_{M}(y)$ in black can be described as a broad, uni-modal PDF that roughly resembles a log-normal distribution characteristic of supersonic turbulence.}
 \label{fig:pdfs_phases}
\end{figure}

\subsection{Halo Gas Properties in the Simulations}

\begin{figure*}
 \centering
 \includegraphics[width=\textwidth]{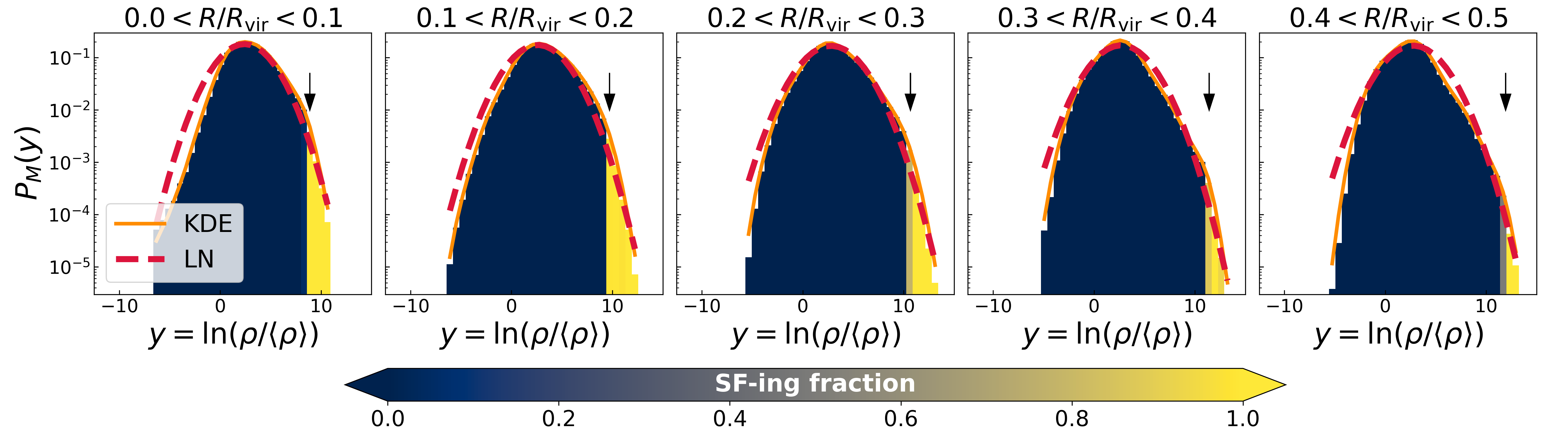}
 \caption{PDFs of the mass-weighted gas density contrast $y=\ln(\rho/\langle \rho \rangle)$ in different radial bins for the simulation $\texttt{z5m11h}$. The histograms show the PDFs directly measured from the simulation after combining the data from all snapshots over $5<z<6$. The histograms are fit by a log-normal function (``LN''; see Equation~(\ref{eq:pdf_ln}); red dashed curve) and a KDE method (orange solid curve). Each PDF is further color-coded by the mass fraction of star-forming gas evaluated from the simulation based on gas particles with $\mathrm{SFR}>0$. For comparison, the downward arrows indicate the $y$ value corresponding to the gas density threshold imposed for star formation in the FIRE-2 simulations ($n_\mathrm{crit}=1000\,\mathrm{cm^{-3}}$). The agreement with the color-coding indicates that the estimate of the star-forming gas mass fraction based on the density PDF and a fixed threshold, as in Equation~(\ref{eq:f_sf}), is an excellent approximation to the simulations.}
 \label{fig:pdfs_fits}
\end{figure*}

In Figure~\ref{fig:maps}, we show projections of the mass-weighted average gas temperature, number density, and line-of-sight velocity of the simulation $\texttt{z5m11h}$ at redshifts $z\sim5$ and 6. The gas distribution is characterized by a highly dynamic and disordered morphology, with large temperature and density fluctuations that spatially extend to large radii \citep[see also][]{Stern2021, Hafen2022}. This is in stark contrast with the gas distribution in $\sim$\,$L^{*}$ SFGs at lower redshift where a stable rotating disk surrounded by a hot CGM is commonly found. The difference is likely due to the short cooling times in the halo which prevents thermalization of inflows and outflows in the halo, coupled with a lack of sufficiently deep and concentrated gravitational potential that prevents the formation of a disk at the center \citep{Hopkins2023}. In such systems, the energy budget of the gas reservoir is strongly dominated by the kinetic energy of turbulence and coherent inflows/outflows, instead of rotation and thermal energy, and there is no clear distinction between the ISM and CGM \citep{Gurvich2023}. Similar to the gas, ongoing star formation (measured by star particles formed in the past 10\,Myr) as marked by the green contours also appears clumpy and spatially extended, generally tracing the distribution of cold and dense gas. By comparing with the locations of identified subhalos, we are able to verify that neither these star-forming clumps nor the cold and dense gas traced by them are correlated with subhalos in any significant way. In Appendix~\ref{sec:appendix-subhalo}, we quantitatively characterize the star formation associated with subhalos in our simulations (see Figures~\ref{fig:subhalo_vis} and \ref{fig:subhalo_frac}). 

\begin{figure}
 \centering
 \includegraphics[width=\columnwidth]{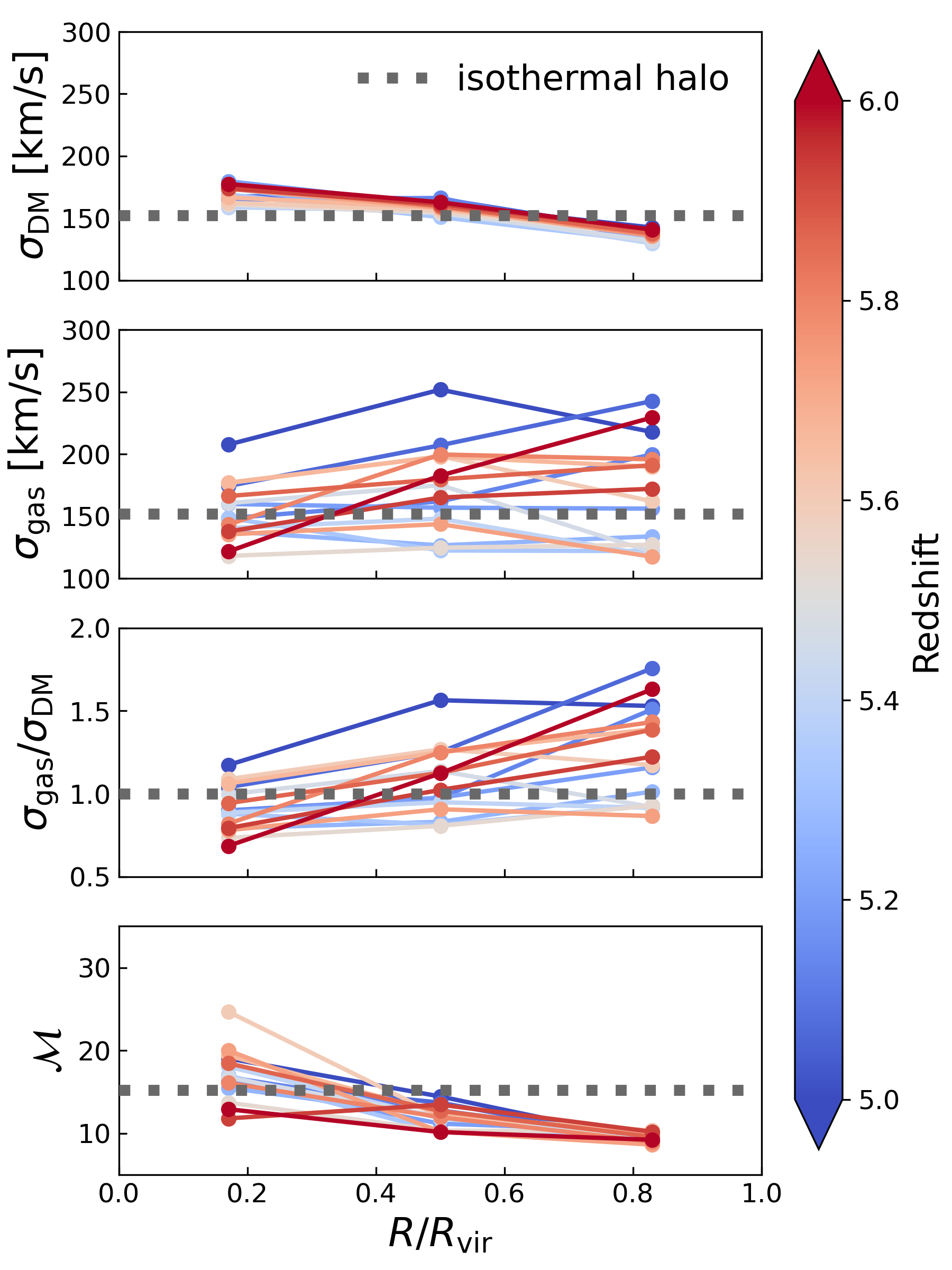}
 \caption{From top to bottom: the dark matter velocity dispersion, the gas velocity dispersion, the ratio of gas and dark matter velocity dispersions, and the Mach number (as defined in Equation~(\ref{eq:mach})) measured in three radial bins. Different colors represent measurements of the 16 snapshots of the simulation $\texttt{z5m11h}$ over $5<z<6$. For comparison, results calculated using our simple analytic model assuming an isothermal halo density profile and a speed of sound $c_\mathrm{s} = 10\,\mathrm{km/s}$ (typical for volume-filling warm neutral/ionized gas) are shown by the horizontal dotted lines.}
 \label{fig:mach_number}
\end{figure}

The dominance of kinetic (velocity dispersion) support over rotational and thermal support for gas across the galaxy host halo is quantified in Figure~\ref{fig:comp_energetics}, where we show a comparison of different forms of gas energy as a function of halo radius. Here and in our subsequent analysis, we consider 10 evenly spaced radial bins with $\Delta R/R_\mathrm{vir}=0.1$ between $R=0$ and $R=R_\mathrm{vir}$. The left panel shows that the total kinetic energy (summed over gas particles of mass $m_i$), 
\begin{equation}
E_\mathrm{kin} = E_{R} + E_{\theta} + E_{\phi} = \sum_{j=R,\theta,\phi} \sum_{i} \frac{1}{2} m_i v_{i,j}^2,
\end{equation}
dominates over the thermal energy, 
\begin{equation}
E_\mathrm{therm} = \sum_{i} \frac{3}{2} \frac{k T_\mathrm{gas}}{\mu m_\mathrm{p}} m_i
\end{equation}
at all radii and the majority of $E_\mathrm{kin}$ is in the form of turbulent energy
\begin{equation}
E_\mathrm{turb} = \sum_{j=R,\theta,\phi} \sum_{i} \frac{1}{2} m_i (v_{i,j}-\bar{v}_{i,j})^2, 
\label{eq:Eturb}
\end{equation}
associated with velocity dispersions. The right panel shows the decomposition of kinetic energy in spherical polar coordinates, with the polar axis defined to align with the total angular momentum of gas within $0.2\,R_\mathrm{vir}$. We note that the rotational energy does not dominate at any radius. Meanwhile, while there appears to be a rough equipartition of $E_{R}$, $E_{\theta}$, and $E_{\phi}$ in the inner halo, at large halo radii $E_{R}$ becomes increasingly significant as a result of inflows/outflows, consistent with the picture where the turbulence is driven by feedback and accretion \citep{Goldner2025}. A similar partitioning of the gas energy is seen for the simulated ISM/inner CGM of Milky-Way progenitors in their early, bursty stage of formation at lower redshifts \citep{Stern2021,Gurvich2023}. 

\begin{figure*}
 \centering
 \includegraphics[width=\textwidth]{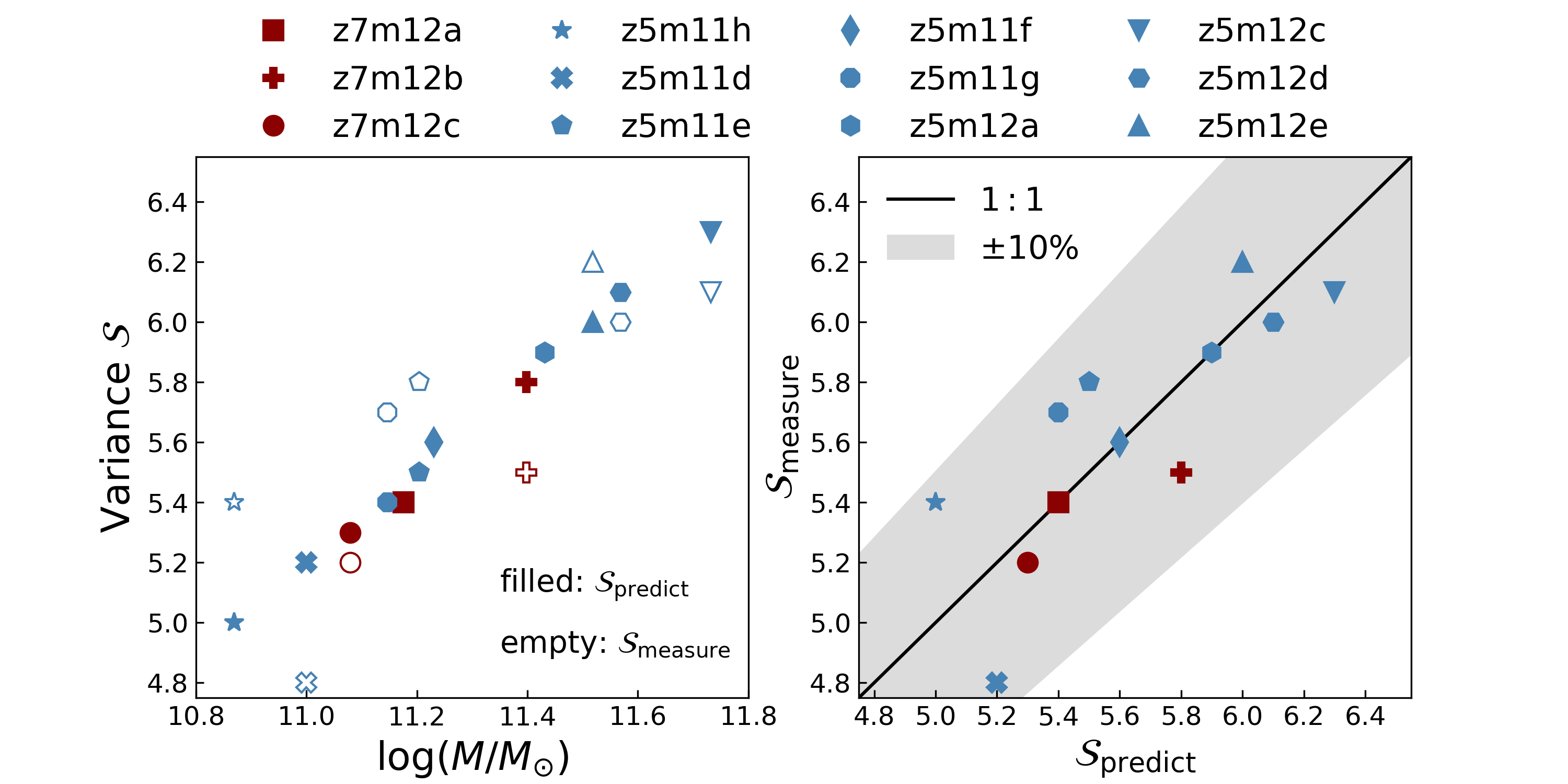}
 \caption{A comparison of the measured and predicted variance, $\mathcal{S}$, of the gas density contrast distribution for the full set of high-$z$ FIRE-2 simulations analyzed in this work (the simulation names are listed in the top legend). Analytic predictions of $\mathcal{S}$ are derived from the halo velocity dispersion corresponding to the halo mass $M_\mathrm{vir}$ assuming $b=0.8$ (see Sections~\ref{sec:model-turbulence:profile} and \ref{sec:model-turbulence:distribution}) and compared against the best-fit $\mathcal{S}$ value assuming a log-normal PDF measured directly from simulations (the median value of measurements in radial bins $<0.5 R_\mathrm{vir}$ is taken). The good agreement between measured and predicted $\mathcal{S}$ at the $\sim10\%$ level over a wide halo mass range confirms the validity of our simple but physically motivated approximation based on our turbulent framework.}
 \label{fig:S_comp}
\end{figure*}

\begin{figure*}
 \centering
 \includegraphics[width=\textwidth]{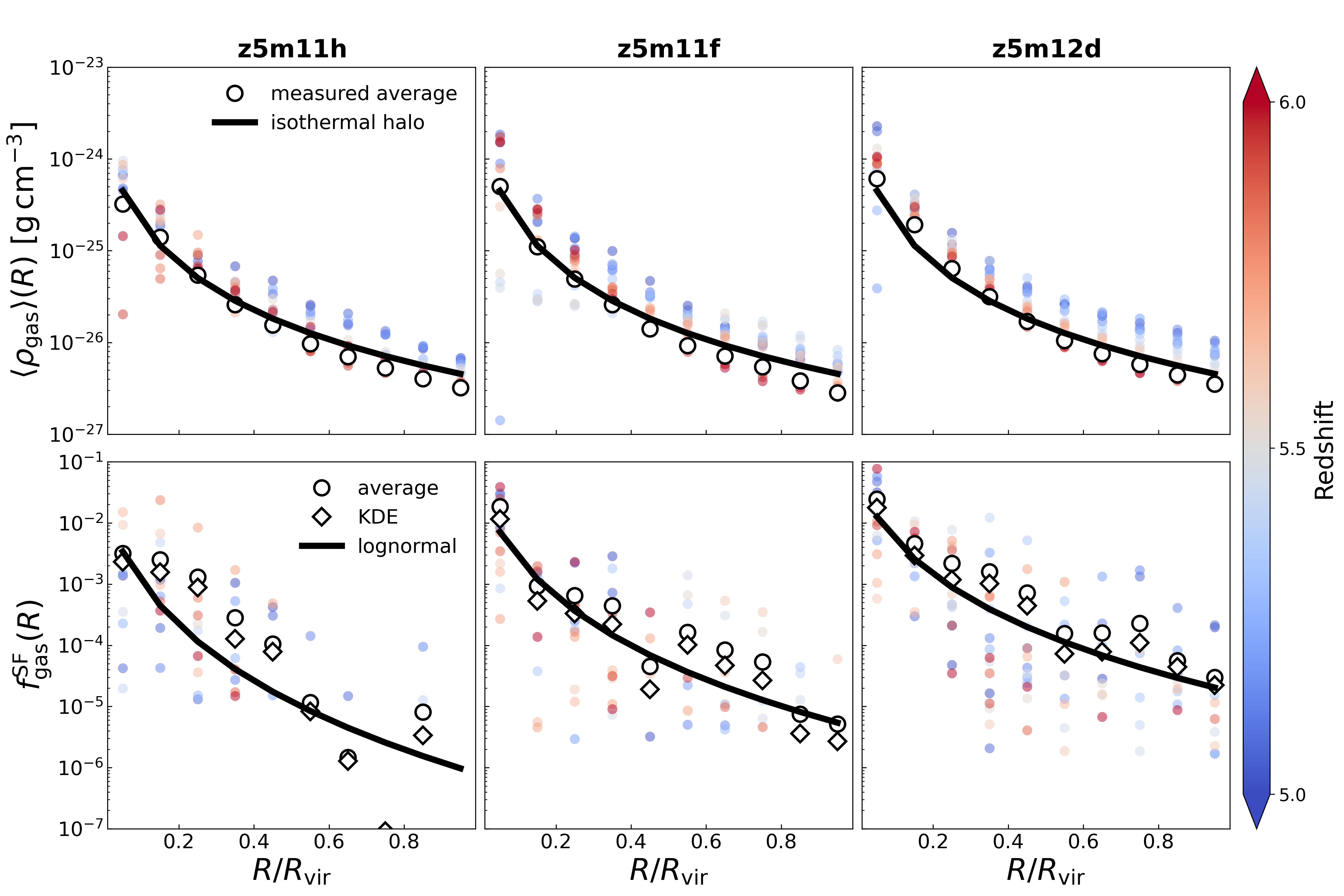}
 \caption{Halo profiles of the volume-weighted gas mass density (top) and star-forming gas mass fraction (bottom) for the three FIRE-2 simulations ($\texttt{z5m11h}$, $\texttt{z5m11f}$, $\texttt{z5m12d}$) considered in this work. Measured values from individual simulation snapshots over $5<z<6$ are shown as small color-coded circles, whereas the average values measured from all snapshots combined are shown as white circles. In the bottom panels, mass fractions evaluated using the measured gas mass density and the integral of the best-fit log-normal density contrast PDF are shown as white diamonds. The solid curves show the predictions of our simple analytic model as specified in Section~\ref{sec:model-turbulence}, which assumes an isothermal halo profile and a log-normal density contrast PDF, with constants $f_\mathrm{gas} = 0.1$ and $b = 0.8$ calibrated to the simulations to reproduce the measured halo gas density profiles and variances of the density PDF.}
 \label{fig:gas_profiles}
\end{figure*}

A fundamental hypothesis in Section~\ref{sec:model-turbulence} to be validated against simulations is that the PDF of the gas density contrast $y$ can be well approximated as one singly-peaked, log-normal-like distribution (or some modified version of it, e.g., with a power-law tail at high densities). In what follows, we focus on the characterization of this density contrast PDF in different radial bins of our simulated galaxies. For the analysis of each radial bin, we combine the particle information from 16 snapshots of the simulated galaxy (saved every $\sim15\,$Myr) between $z=5$ and 6 to derive the statistics of interest in a way that is less susceptible to temporal variations caused by the stochasticity of e.g., stellar feedback and gas inflows\footnote{As shown in Figure~\ref{fig:S_comp}, three additional simulations are analyzed at higher redshift ($8<z<9$) using the same method.}. Unless otherwise color-coded by redshift (e.g., in Figure~\ref{fig:gas_profiles}), we consider the redshift-averaged $\langle \rho \rangle$ for the PDF analysis. We note that while the total averaging time interval (250\,Myr) is a few times the halo dynamical timescale $t_\mathrm{dyn}\sim0.1/H(z)$ at these redshifts, the time variability and halo-to-halo variations are not expected to completely averaged out and may therefore contribute in part to the discrepancy with our simple analytic model. 

Figure~\ref{fig:pdfs_phases} shows the normalized PDFs of $y$ evaluated for cold ($T_\mathrm{gas}<10^3\,$K), warm ($10^3<T_\mathrm{gas}<10^5\,$K), and hot ($T_\mathrm{gas}>10^5\,$K) gas at inner ($0<R/R_\mathrm{vir}<0.1$) and outer halo radii ($0.5<R/R_\mathrm{vir}<0.6$). From how the total PDF is composed of individual PDFs of gas in different temperature ranges, it is evident that the total $y$ distribution can be described by a singly-peaked PDF despite its multiple components corresponding to different gas phases. Here, the turbulent gas is not segregated into discrete phases that lead to a multi-modal distribution with well separated peaks. This is likely as a result of the dominance of kinetic pressure $P_\mathrm{turb}$ over thermal pressure $P_\mathrm{th} = n k_\mathrm{B} T_\mathrm{gas}$ due to the short cooling times of the hot gas (Figure~\ref{fig:comp_energetics}), which implies that ram pressure dominates over thermal pressure in setting the density distribution \citep{Kakoly2025}. From Figure~\ref{fig:pdfs_phases}, the warm gas contributes the most to the total PDF and largely sets its peak location, whereas the cold (hot) gas with on average higher (lower) density contrast effectively broadens the total PDF. 

To understand what functional form can appropriately describe the $y$ distribution, we examine the log-normal parameterization of the density PDF specified in Section~\ref{sec:model-turbulence:distribution} by applying a least square fitting to the $y$ distribution measured from the simulations and then compare the best-fit results with the PDF obtained from kernel density estimation (KDE). In Figure~\ref{fig:pdfs_fits}, we use the simulation $\texttt{z5m11h}$ as an example to illustrate how well the simulated $y$ distribution is captured by a log-normal distribution (Equation~(\ref{eq:pdf_ln})) and the KDE respectively in different radial bins. Compared to the histogram directly sampled from the simulation snapshots and its KDE approximation, Equations~(\ref{eq:pdf_ln}) provides a reasonably good fit to the measured distribution in the inner half halo where most of the spatially extended star formation occurs (see Figure~\ref{fig:maps}). For comparison, in Appendix~\ref{sec:appendix-ln+pl} we examine an alternative case where the PDF is given by a piecewise distribution composed of a log-normal function with a power-law tail at high densities. 

Apart from comparing functional forms that describe the $y$ distribution, we also color-code the histograms in Figure~\ref{fig:pdfs_fits} according to the mass fraction of star-forming gas evaluated from the simulations by summing up the gas particles whose $\mathrm{SFR}$ is greater than 0. Since the FIRE-2 simulations impose a gas number density threshold $n_\mathrm{crit}=1000\,\mathrm{cm^{-3}}$ for star formation to happen, as indicated by the downward arrow, this allows us to verify whether it is reasonable to model star formation from the turbulent gas density field using the density PDF and some density threshold as specified in Equation~(\ref{eq:f_sf}). The color-coding suggests a sharp transition of the star-forming gas mass fraction from 0 to 1 once the critical density assumed (labeled by the downward arrow) is surpassed. The sharp dichotomy between non-star-forming and star-forming gas here is a consequence the critical density $n_\mathrm{crit}$, which prevents star formation in low-density gas and is more restrictive than other FIRE-2 star formation criteria (self-gravitating, self-shielding, and Jeans-unstable) in the regime studied (see Appendix C of \citealt{Hopkins2018}). Since the gas density profile decreases with the radial distance, the transition occurs at a higher density contrast at a larger halo radius, but the close agreement between the assumed $n_\mathrm{crit}$ and where the transition is measured persists. This strongly motivates the idea outlined in Section~\ref{sec:model-turbulence:sf} to derive star formation rate profiles from reasonable approximations of the turbulent density field and star formation criteria. 

\begin{figure*}
 \centering
 \includegraphics[width=\textwidth]{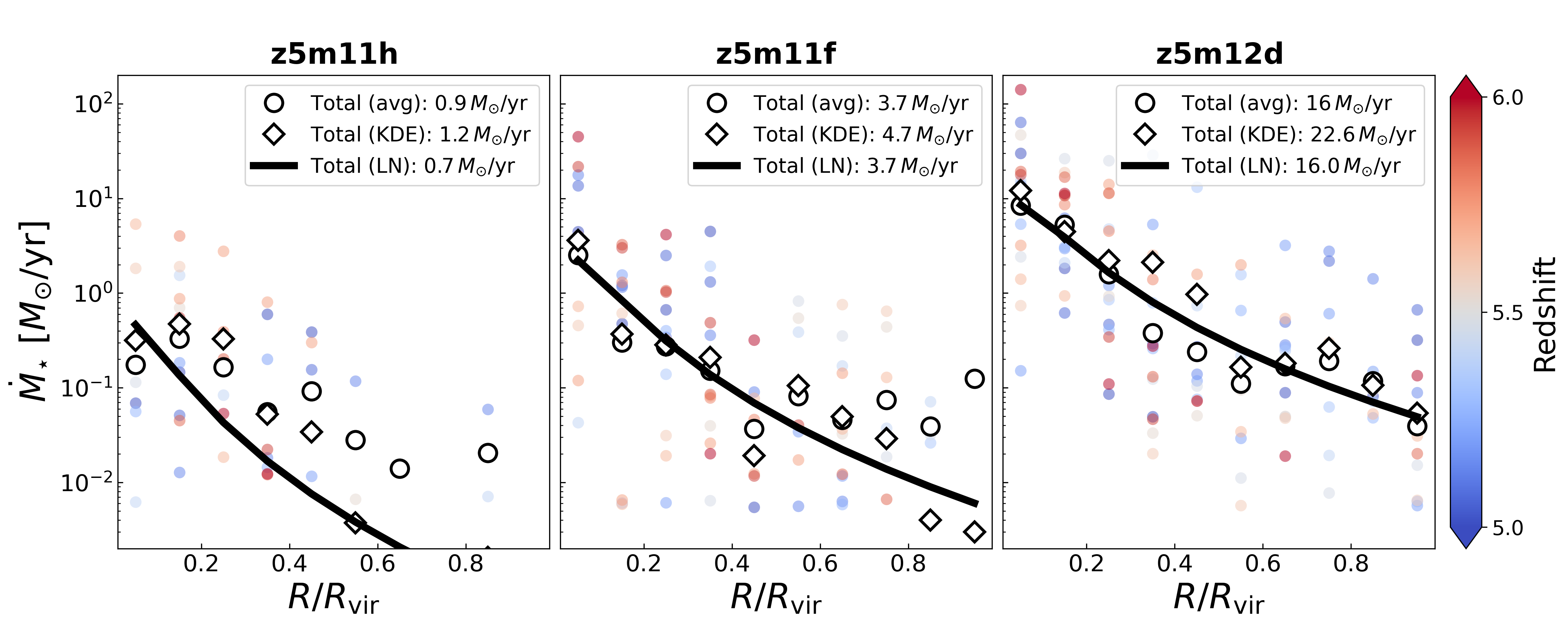}
 \caption{Same as Figure~\ref{fig:gas_profiles} but for the radial profile of the SFR in the halo. The circles denote the total SFR of gas particles per radial bin at each snapshot (colored) or averaged over $5<z<6$ (white). The turbulent and bursty nature of these galaxies leads to order-of-magnitude time variations of these SFR profiles. The diamonds and the solid curve show predictions of our analytic approximation using equation~(\ref{eq:sfrgrad}) for the KDE and log-normal fits to the density PDF, respectively. Given the large scatter, the time-averaged SFR profile is well captured by our simple analytic model that assumes a log-normal PDF. The total SFR summed over all radial bins is quoted in the legend for each measure of the SFR.}
 \label{fig:sfr_profiles}
\end{figure*}

As introduced in Section~\ref{sec:model-turbulence:distribution}, the variance of the log-normal distribution, $\mathcal{S}$, is a key quantity to model based on the turbulence properties in order to predict the gas density distribution and therefore the mass fraction of gas that can form stars across the galaxy host halo. Figure~\ref{fig:mach_number} shows comparisons between the velocity dispersions and Mach numbers measured from the simulations and those predicted by our simple analytic model (Equations~(\ref{eq:S}) and (\ref{eq:mach})). These comparisons suggest that $\sigma_{v,\mathrm{gas}} \approx \sigma_{v}(M_\mathrm{vir},z)$ is a reasonably good assumption, though it might be oversimplified to assume a fixed $c_\mathrm{s}$ because the higher gas kinetic temperature at large halo radii can result in a higher $c_\mathrm{s}$ and thus lower $\mathcal{M}$. 

In Figure~\ref{fig:S_comp}, we show comparisons between the analytically predicted $\mathcal{S}_\mathrm{predict}$ that can be derived from the Mach number using the halo mass-dependent velocity dispersion (Equation~(\ref{eq:mach})) and the best-fit value $\mathcal{S}_\mathrm{measure}$ directly measured from simulations assuming a log-normal density contrast PDF. We remind that a choice of $b = 0.8$ (motivated by values of $\mathcal{S}_\mathrm{measure}$ of our simulated galaxies) is made when evaluating $\mathcal{S}_\mathrm{predict}$ since it is challenging to analytically predict $b$. The remarkable similarity between the predicted and measured $\mathcal{S}$ values over a wide range of halo mass implies that even though the physics involved in shaping the actual gas density distribution is highly complex, the assumptions made in our simple analytic model are generally reasonable and lead to a decent approximation to the actual gas density contrast PDFs in the FIRE-2 simulations.

In Figure~\ref{fig:gas_profiles}, we show halo radial profiles of the volume-weighted average gas mass density, $\langle \rho_\mathrm{gas} \rangle$, and the star-forming gas mass fraction, $f^\mathrm{SF}_\mathrm{gas}$, for the simulations $\texttt{z5m11h}$, $\texttt{z5m11f}$, and $\texttt{z5m12d}$, respectively. For comparison, measurements made for the 16 individual snapshots over $5<z<6$ are shown in the background by the small circles in color, whereas the average values from all 16 snapshots combined are shown by the larger white circles. For $f^\mathrm{SF}_\mathrm{gas}$, we also include simulation-based results obtained using measured $\langle \rho_\mathrm{gas} \rangle$ profiles and the integral of the density contrast PDF approximated by the KDE method, $\int_{>y_\mathrm{crit}} P_{M,\mathrm{KDE}}(y) d y$. These results are displayed as white diamonds to be compared against the white circles based on counting gas particles that are forming stars (i.e., with an instantaneous $\mathrm{SFR}>0$). Predictions derived from the assumptions made in our simple analytic model in Section~\ref{sec:model-turbulence} are shown by the solid curves, which are evaluated using $M_\mathrm{vir}$ of the simulated galaxies at $z=5.5$. While in simulations significant time variations are present for these profiles as indicated by the circles in different colors and expected from stochasticity in, e.g., feedback, gas inflows, and mergers captured in the simulations, their time averages shown by the white circles are in general agreement with our simple analytic approximation. This confirms that a combination of a log-normal function and a constant density threshold can reasonably well approximate the mass fraction of star-forming gas in the simulated galaxies out to large halo radii. 

We note again that the plotted profiles of $\langle \rho_\mathrm{gas} \rangle$ and $f^\mathrm{SF}_\mathrm{gas}$ from our analytic model assume a fixed gas fraction $f_\mathrm{gas}=0.1$ and a softening parameter $r_\mathrm{s} = 0.05 R_\mathrm{vir}$, both of which are chosen to provide good approximations of the time-averaged results from simulations. The physics of, e.g., inflows and feedback-driven outflows that can regulate the amplitude of $f_\mathrm{gas}$ and its dependence on halo mass and redshift, are outside the scope of this model and must either be assumed or modeled separately. Nevertheless, these comparisons show encouraging evidence for our hypothesis that it is possible to generalize the turbulent framework initially developed for molecular clouds to describe the halo-scale gas properties of the highly dynamic, low-mass SFGs that prevail in the high-$z$ universe. It is then interesting to check if the halo profile of star formation properties is also well captured by this turbulent picture. 

\subsection{Connecting Gas Properties to Star Formation}

With the gas mean density profile and the density contrast PDF in hand, we can verify how closely a simple star formation prescription like Equation~(\ref{eq:sfrgrad}) captures the spatial distribution of star formation on halo scales for our simulated galaxies. In Figure~\ref{fig:sfr_profiles}, we show radial profiles of the SFR for the same simulations shown in Figure~\ref{fig:gas_profiles}. At a given radius, SFRs of gas particles in the simulations are used to measure the SFRs for individual snapshots or the average with all 16 snapshots combined. On the other hand, we use the analytic prescriptions in Section~\ref{sec:model-turbulence:sf} to predict the SFR profile from the gas mass gradient and $f^\mathrm{SF}_\mathrm{gas}$, assuming $\epsilon_\mathrm{SF}=100\%$ and evaluating $t_\mathrm{ff}$ at $n_\mathrm{crit}=10^3\,\mathrm{cm}^{-3}$, consistent with the assumptions made by in the FIRE-2 simulations \citep{Hopkins2018,Ma2018Morph,Ma2018LF}. We emphasize that our analytic approximations rely on only the halo mass as the input variable and assume fixed values of $f_\mathrm{gas} (= 0.1)$ and $b (= 0.8)$ consistent with the simulations. As shown by the comparison of the simulation average, the KDE approximation, and our analytic model based on the best-fit log-normal distribution to $P_{M}(y)$, the simulated SFR radial profile is well reproduced in our turbulent framework, including for the two more massive galaxies \texttt{z5m11f} and \texttt{z5m12d} whose (time-averaged) SFR profiles extend beyond $0.5R_\mathrm{vir}$. 

We calculate and label in the legend the total SFR summed over all radial bins for each simulation, which is mainly set by the contribution from inner radii. Notably, the average total SFR measured over $5<z<6$ agrees with the prediction of our analytic approximation (describing the density contrast PDF with either the KDE or the best-fit log-normal distribution) to better than a few tens percent for all three galaxies. Given the large scatter in the SFR due to spatial and temporal variations of star formation in any radial bin across snapshots, both the SFR profile and the total halo SFR predicted by our turbulent framework are in good agreement with the simulations. These comparisons provide compelling evidence for the validity of using a halo-scale turbulent flow to model the star formation properties across high-$z$ SFGs found in the FIRE-2 simulations. 

\begin{figure}
 \centering
 \includegraphics[width=\columnwidth]{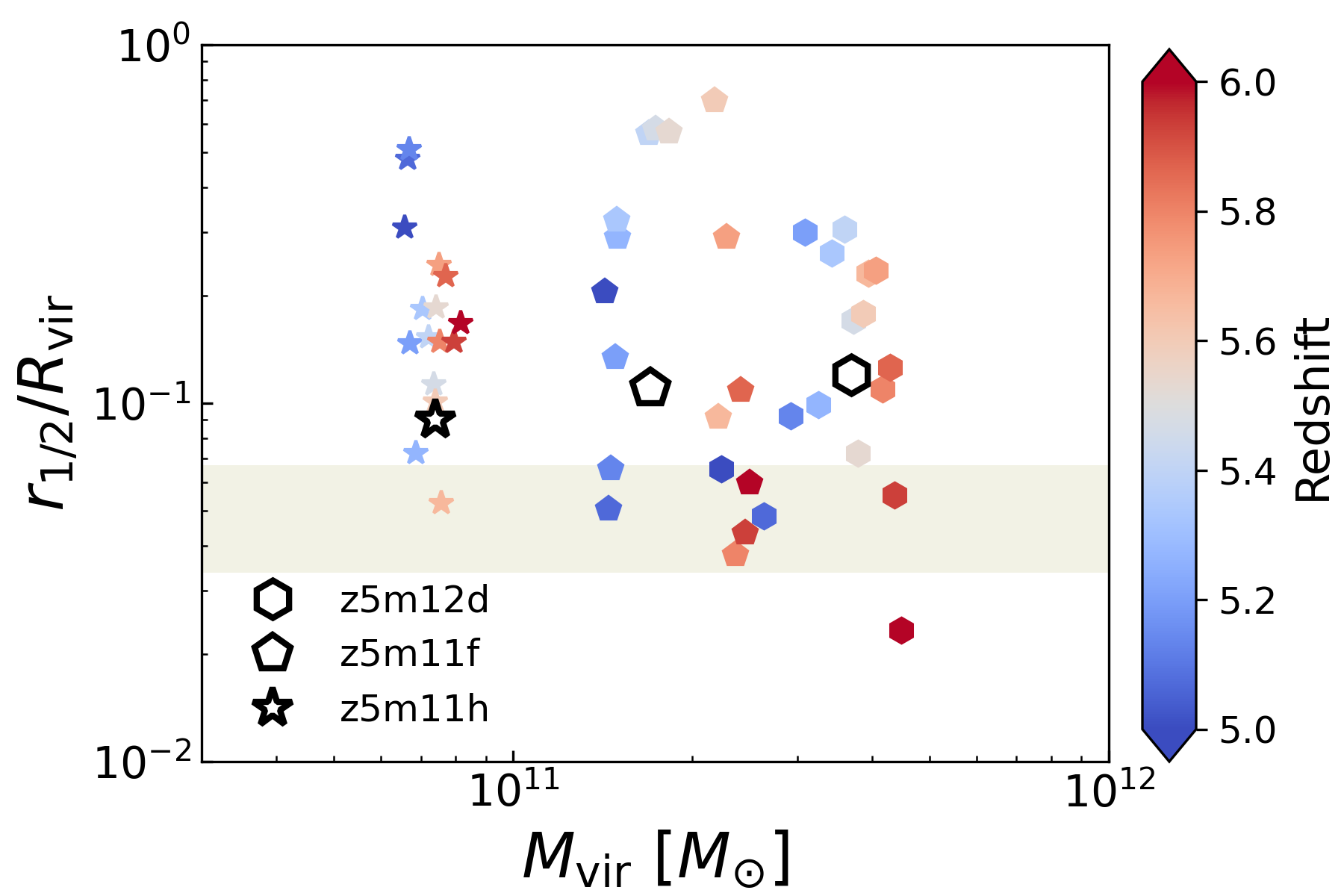}
 \caption{Comparison of the galaxy half-SFR radius, $r_{1/2}$, measured from three simulations (small color-coded symbols) and predictions by our simple analytic model (large white symbols). For reference, the shaded band shows the rough expectation for an exponential disk often considered in the literature, assuming a spin parameter $\lambda \sim 0.02$--0.04 appropriate for high-$z$ halos \citep{Yung2024DM}. The bursty high-$z$ SFGs in the simulations have sizes well approximated on average by our turbulent model, but are generally larger than the exponential disk model predictions.}
 \label{fig:size}
\end{figure}

With the SFR profile in hand, we can further measure galaxy sizes and compare them with the simulation results. Specifically, we adopt the half-SFR radius, $r_{1/2}$, as a proxy for the UV half-light radius of galaxies, as commonly reported in observations \citep{Abdurrouf2023,Bassini2023}. In Figure~\ref{fig:size}, we plot $r_{1/2}$ (defined with respect to the total SFR within $R_\mathrm{vir}$) measured from the simulation snapshots in color against the predictions of our analytic model (evaluated using $M_\mathrm{vir}$ of the simulated galaxy at $z=5.5$) for the three example galaxies shown above. For comparison, we also show the expected galaxy sizes from the commonly assumed exponential disk formation model that suggests $r_{1/2} \simeq 1.68 r_\mathrm{d} \simeq 1.68 \lambda R_\mathrm{vir}$, where $r_\mathrm{d}$ is the scale radius and $\lambda$ is the halo spin parameter associated with large-scale gravitational torques \citep{Mo1998}. Here, the shaded band corresponds to $\lambda = 0.02$--0.04, a plausible range for high-$z$ halos that agrees with the predictions of $N$-body simulations \citep{Danovich2015,Yung2024DM}. Despite the significant scatter due to burst cycles of star formation, it is evident from both the direct simulation measurements and our analytic turbulent model that the FIRE-simulated high-$z$ SFGs on average have larger sizes (comparable to 10\% of $R_\mathrm{vir}$) than what the exponential disk model predicts. These results are consistent with the spatially extended star formation illustrated in Figure~\ref{fig:maps}, a key property of the high-$z$ turbulence-dominated halos with large density fluctuations that extend to large halo radii.  

\section{Implications for the High-$z$ SFE} \label{sec:implications}

The SFR predicted by our analytic framework is useful for understanding the galaxy-scale SFE of bursty SFGs in the high-$z$ universe. Recent JWST observations have revealed evidence of a high number density of UV-bright galaxies at $z\gtrsim10$ that exceeds the predictions of many semi-analytic/empirical models calibrated to pre-JWST observations (e.g., \citealt{Mason2015}; \citealt{SF2016}; \citealt{Tacchella2018}; \citealt{Yung2019}; \citealt{Harikane2022}; though note that the FIRE-2 simulations predict an abundance of bright galaxies in better agreement with the JWST observations, as shown by the UV luminosity functions predicted in \citealt{Ma2018LF,Ma2019}; \citealt{Sun2023}; \citealt{Feldmann2025}). Explaining these observations requires an effectively larger than usual light-to-mass ratio of galaxies, an intrinsically higher SFE per unit baryon available, or a combination of both. 

Several plausible explanations have been proposed and investigated within the standard $\Lambda$CDM framework, including increased SFEs due to inefficient feedback \citep[e.g.,][]{Dekel2023,Li2024}, bursty SFHs \citep[e.g.,][]{Shen2023,Sun2023,Gelli2024,Kravtsov2024}, top-heavy stellar initial mass functions \citep[e.g.,][]{Inayoshi2022,Yung2024}, unrecognized AGN contributions \citep[e.g.,][]{Hegde2024,Trinca2024,Hutter2025}, and evolving dust attenuation associated with the clearance of dust by outflows \citep[e.g.,][]{Ferrara2023,Ferrara2024}. An interesting scenario is one in which the galaxy-scale SFE becomes as high as unity due to strongly suppressed stellar feedback in dense and low-metallicity environments, as suggested by the feedback-free starburst (FFB) model \citep{Dekel2023,Li2024}. Given the profound importance of the SFE for understanding the high-$z$ universe \citep{Shen2024,Boylan-Kolchin2025}, it is valuable to examine how the predictions of our analytic turbulent model, which closely match galaxies in the FIRE simulations, fit within the broader theoretical landscape of the SFE of high-$z$ galaxies.

\begin{figure*}
 \centering
 \includegraphics[width=0.85\textwidth]{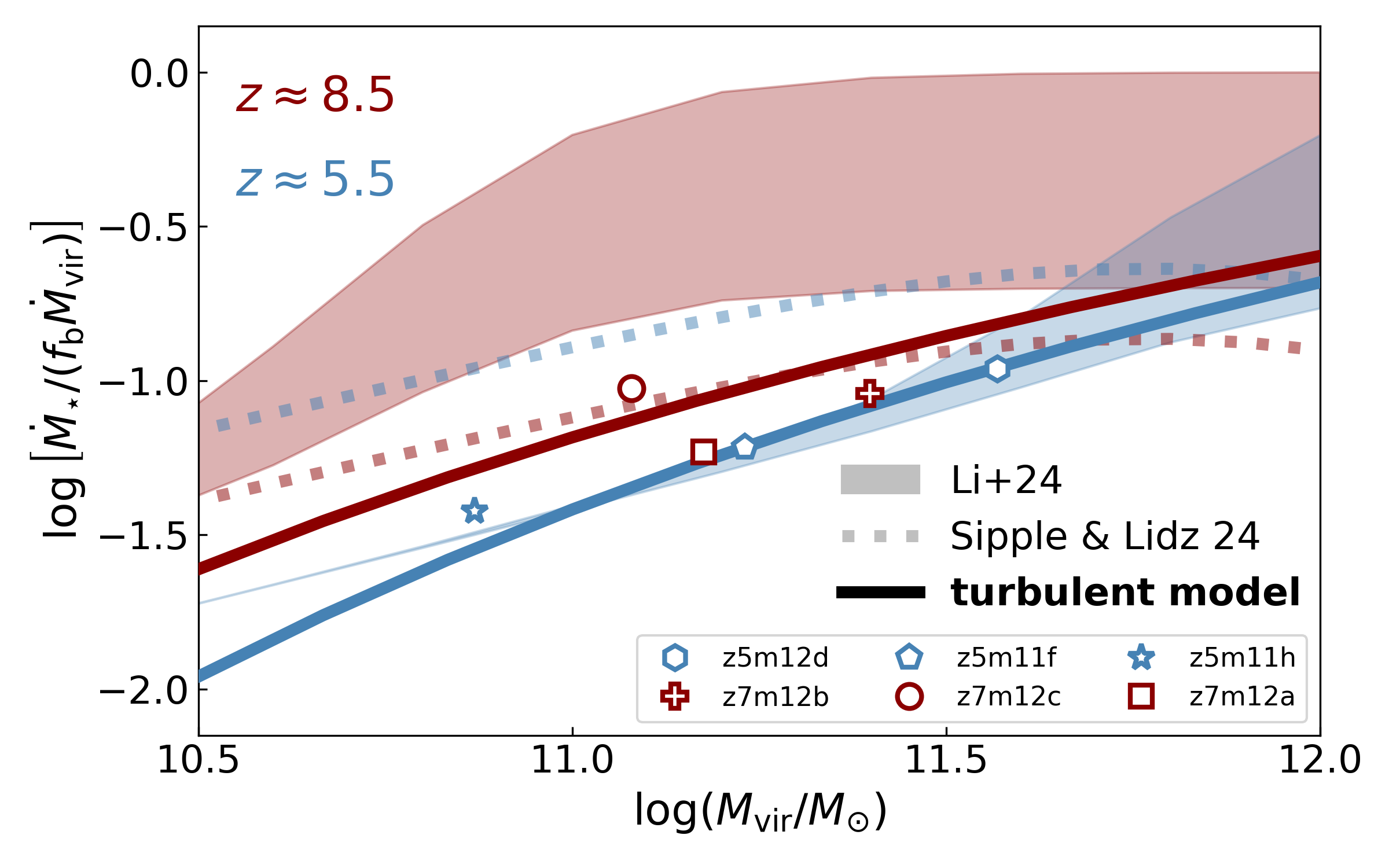}
 \caption{Comparison of the instantaneous, galaxy-scale SFE predicted by our turbulent model (solid curves) as a function of halo mass and the direct measurements from six example FIRE simulated galaxies (averaged over a redshift interval of $\Delta z = 1$) shown by the different symbols at $z\approx5.5$ and 8.5. Also shown are the results derived from halo abundance matching \citep[dotted lines;][]{SL2024} and the feedback-free starburst (FFB) model \citep[shaded bands whose width denotes the range corresponding to a maximum FFB SFE of $0.2 < \epsilon_\mathrm{max} < 1$ as defined in][]{Li2024}. At high redshift (e.g., $z\approx8.5$), unlike the FFB model that predicts increasingly enhanced SFEs in massive halos, our turbulent model (where a high SFE per free-fall time occurs only in dense gas comprising a small fraction of the total $M_\mathrm{gas}$) predicts SFEs noticeably lower and more comparable to the values inferred from abundance matching. A constant $f_\mathrm{gas} = 0.1$ is assumed for our turbulent model predictions (see discussion in the main text).}
 \label{fig:sfe}
\end{figure*}

In Figure~\ref{fig:sfe}, we show the instantaneous galaxy-scale SFE, namely the ratio of the SFR to the baryonic mass accretion rate $\dot{M}_{\star}/(f_\mathrm{b}\dot{M}_\mathrm{h})$, of six FIRE simulated galaxies predicted by our turbulent framework given their halo masses at $z=5.5$ or 8.5, as well as how it compares against the results from the FFB model \citep[shaded bands, from][]{Li2024} or abundance matching the UV luminosity function to the halo mass function \citep[dotted lines, from][]{SL2024} at the same redshifts. Interestingly, while at $z \approx 5.5$ the predictions of our turbulent framework and the FFB model agree very well, our simple model implies a significantly weaker redshift dependence when the predictions for $z \approx 5.5$ and 8.5 are compared. Analytic approximations to the FIRE simulations based on our turbulent framework imply an instantaneous galaxy-scale SFE $\lesssim10\%$. Similarly, results based on abundance matching also show no evidence for a substantially increased, close-to-unity SFE at these redshifts \citep[see also e.g.,][]{Donnan2025}. 

The main difference between the FFB scenario and our model is that in the former case the baryonic (gas) content of the entire galaxy host halo is assumed to form stars at an elevated, and potentially unity, SFE once the feedback-free conditions are met, whereas as demonstrated in Figure~\ref{fig:efficiency_vs_density} our model suggests that the broad gas density distribution resulting from the highly turbulence-dominated halo makes the mass fraction of dense, efficiently star-forming gas very small (up to several percent in the inner halo). It should be emphasized that even though our turbulent framework does not predict a uniformly enhanced SFE across the galaxy host halo, as assumed in the FFB model through a constant conversion factor of the baryonic mass accreted, FFB-like events are implicitly allowed in cold, dense gas where the SFE can be of order unity when evaluated over specific spatial and temporal scales in cold, dense gas. In fact, this occurs at the star formation density threshold in the default FIRE-2 model. 

This can be related to several recent theoretical studies that focus on the physics behind the SFE of high-$z$ galaxies. The density-specific nature of the SFE gives rise to its multi-scale characteristics 
(see e.g., \citealt{Andalman2024}, \citealt{Shen2025}, and \citealt{Wang2025} for investigations based on the RAMSES and THESAN-Zoom simulations), which are central to understanding both the connection and the distinction between star formation properties on halo versus gas cloud scales. As suggested by the turbulent framework presented, variations in the galaxy-scale SFE with mass and redshift are governed by physical processes that shape the detailed gas density distribution, particularly its high-density tail. Indeed, the critical role of the dense gas cloud mass fraction in modulating the galaxy-scale SFE of the galaxies in the high-$z$ universe was also recently studied in semi-analytic models \citep{Somerville2025}, which suggest that the strong dependence of the cloud-scale SFE on gas surface density \citep[see e.g.,][]{Grudic2018,Grudic2020,Menon2022} allows the galaxy-scale SFE to grow with an increasing mass fraction of dense gas clouds. As implied by our turbulent framework, the mass fraction of cold, dense gas emerges as a key physical quantity governed by properties of the turbulent density field across the galaxy's host halo. Understanding the generation and maintenance of turbulent gas can therefore be crucial for informing parameter choices in such semi-analytic models of high-$z$ galaxies. 

\section{Summary} \label{sec:conclusions}

In this paper, we present a novel turbulent framework for understanding and theoretically describing star formation in early, bursty SFGs, in contrast to the classical picture where star formation is assumed to occur within a rotationally supported disk in a self-regulated, quasi-equilibrium state. This framework, motivated by findings from both the latest observations and state-of-the-art simulations of high-$z$ galaxy formation, captures the highly disturbed and velocity dispersion-dominated nature of SFGs in their early stage of formation. It is analogous to the turbulent description of star formation in molecular clouds, but extended to the scales of the galaxy's host halo (Figure~\ref{fig:sketch}). 

We describe halo-scale star formation properties by treating halo gas as a density field of supersonic turbulence driven by processes including stellar feedback and accretion \citep{Goldner2025}. This turbulence avoids thermalization due to the short cooling times, and exhibits wide lognormal density distributions as shown by \cite{Kakoly2025}. Specifically, to determine the amount of star formation, we apply a density threshold for star formation to a log-normal-like gas density contrast distribution, whose variance is determined by the turbulent velocity dispersion anchored to the halo's gravitational potential. Remarkably, when compared against high-$z$ SFGs in the FIRE simulations, this turbulent framework successfully reproduces the spatial distributions of various key physical properties of the turbulent gas, especially the amount of cold, dense gas available to form stars. This allows us to apply this turbulent framework to reasonably approximate the time-averaged profile of star formation activities in high-$z$ SFGs. 

By modeling star formation across the halo with turbulence-driven gas density distributions, our framework highlights the importance of turbulence for unraveling the underlying physics governing star formation in high-$z$ galaxies. In particular, it provides a new analytic perspective into modeling high-$z$ SFGs distinct from the classical picture based on the existence and maintenance of a self-regulated, quasi-equilibrium disk. We note that the descriptive power of this analytic model could be enhanced by incorporating additional physical prescriptions in the future. The fact that parameters such as the gas fraction of halos $f_\mathrm{gas}$ (Equation~(\ref{eq:tsip})) and the turbulence parameter $b$ (Equation~(\ref{eq:mach})) are treated as external inputs and calibrated to match the FIRE simulations adds flexibility to our framework. Modeling turbulence in further detail could illuminate how modifications of the density PDF impact the predictions of star formation and other related processes. Similarly, the potential time and mass dependence of $f_\mathrm{gas}$ could be more accurately accounted for by integrating our framework with gas regulator-like models for a more comprehensive description of the galaxy formation physics. In addition, it would be interesting to experiment with more sophisticated star formation criteria than the simple density threshold considered in this paper, such as based on self gravity. Incorporating physical mechanisms that drive the time variability of gas and star formation properties may also be valuable for extending the model’s predictive power beyond time-averaged quantities \citep{FurlanettoMirocha2022,Pallottini2025}. 

While our primary focus in this work is on SFGs in the high-$z$ universe, the turbulent framework presented could be generalized to predict and interpret star formation in low-mass, bursty galaxies at lower redshifts that have yet to settle into disks and remain similarly turbulence-dominated \citep{Stern2021,Gurvich2023}. These low-mass galaxies are important building blocks or progenitors of more massive galaxies such as the Milky Way, and like high-$z$ bursty galaxies, they are not adequately described as equilibrium disks. We note that the halo-scale turbulence discussed here implies a physical alternative to mergers and violent disk instabilities \citep[e.g.,][]{2009ApJ...703..785D} for the origin of star-forming clumps in low-mass galaxies, studied most recently using JWST \citep[e.g.,][]{delaVega2025, Sok2025}. 
\cite{Mandelker2025} analyzed the role of compressive turbulence in clump formation, but focused on gas-rich disks. The lack of significant rotational support in the turbulence-dominated regime distinguishes this limit from what may be described by turbulence-modified disk instability criteria \citep{Romeo2010,RomeoAgertz2014}. Meanwhile, the implications of the turbulence-dominated picture for high-$z$ galaxy host halos extend beyond the star formation properties examined in this work. Valuable insights may be drawn from the halo-scale turbulent density field into the distribution of circumgalactic neutral gas that absorbs ionizing photons \citep{Stern2021DLA}, circumgalactic UV absorption signatures by metals \citep{Kakoly2025}, and the spatial extent of dust in and around high-$z$ galaxies \citep{ZhaoFurlanetto2024}. Building on the success of the presented framework in capturing key aspects of high-$z$ SFGs, future extensions would provide deeper insights for understanding a broad range of theoretical properties and observational signatures of high-$z$ galaxy formation. \\

The authors thank the anonymous referee for helpful comments that improved this paper, as well as Alexander de la Vega, Steven Furlanetto, Aharon Kakoly, Nir Mandelker, Marta Reina-Campos, and Rachel Somerville for stimulating discussion. GS was supported by a CIERA Postdoctoral Fellowship, and would like to acknowledge the Kavli Institute for Theoretical Physics (KITP) where part of this work was done for their hospitality (supported in part by grant NSF PHY-2309135). CAFG was supported by NSF through grants AST-2108230 and AST-2307327; by NASA through grants 21-ATP21-0036 and 23-ATP23-0008; and by STScI through grant JWST-AR-03252.001-A. This work was performed in part at Aspen Center for Physics, which is supported by National Science Foundation grant PHY-2210452. JS was supported by the Israel Science Foundation (grant No. 2584/21). The simulations used in this paper were run on ACCESS computational resources (allocation TG-AST21010). Additional analysis was done using the Quest computing cluster at Northwestern University. 

\software{\textsc{GizmoAnalysis} \citep{Wetzel_2016,2020ascl.soft02015W}, matplotlib \citep{Hunter2007}, 
numpy \citep{Harris2020}, scipy \citep{Virtanen2020}
}

\appendix
\twocolumngrid

\begin{figure*}
 \centering
 \includegraphics[width=\textwidth]{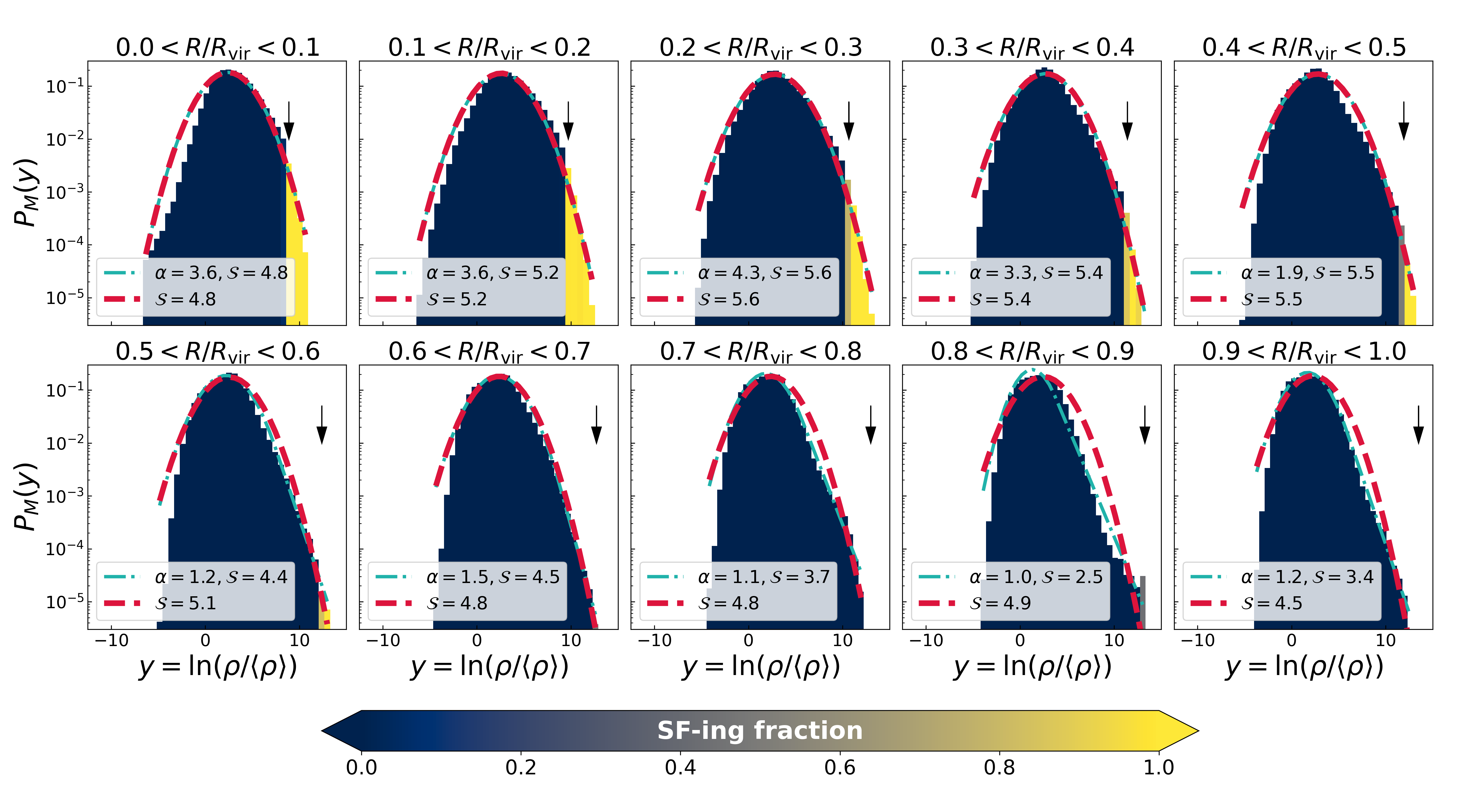}
 \caption{Similar to Figure~\ref{fig:pdfs_fits}, but for a comparison between the pure log-normal and the log-normal+power-law fits (see Equation~(\ref{eq:pdf_piecewise})). Compared with the pure log-normal fit that involves only the variance $\mathcal{S}$, the log-normal+power-law fit involving an extra parameter $\alpha$ provides a better description of the PDF only in the outer half halo (where little star formation occurs), as indicated by the different best-fit values of $\mathcal{S}$ from the two fits.}
 \label{fig:pdfs_fits_more}
\end{figure*}

\section{Log-normal + power-law fits to the density contrast PDF} \label{sec:appendix-ln+pl}

In addition to the log-normal PDF as the simplest function form that describes the density distribution, we further consider a slightly more sophisticated description discussed in \citet{BurkhartMocz2019} for a gravo-turbulent medium, where the PDF consists of a log-normal component at low-to-moderate densities and power-law tail at high densities due to self-gravity of the gas. Specifically, the transition point occurs at a density contrast $y_\mathrm{t} = \ln(\rho_\mathrm{gas,t}/\langle \rho_\mathrm{gas} \rangle)$, which relates to the variance of the log-normal distribution, $\mathcal{S}$, by a single parameter $\alpha$ (the slope of the power-law tail), namely
\begin{equation}
y_\mathrm{t} = (\alpha + 0.5) \mathcal{S}. 
\label{eq:y_trans}
\end{equation}
Equation~(\ref{eq:y_trans}) follows the assumption that the log-normal-to-power-law transition should be smooth. The piecewise parameterization of $P_{M}(y)$ is \citep{Collins2012,Burkhart2018,BurkhartMocz2019},
\begin{align}
P_{M}(y) \mathrm d y = 
  \begin{cases}
    \frac{N}{\sqrt{2\pi \mathcal{S}}} \exp \left[ -\frac{(y - y_0)^2}{2\mathcal{S}} \right] \mathrm d y, & \text{$y < y_\mathrm{t}$} \\
    N C \exp(-\alpha y), & \text{$y \geq y_\mathrm{t}$}
  \end{cases}
\label{eq:pdf_piecewise}
\end{align}
where
\begin{equation}
N = \left[ \frac{C \exp(-\alpha y_\mathrm{t})}{\alpha} + \frac{1}{2} + \frac{1}{2} \mathrm{erfc}\left( \frac{2 y_\mathrm{t} - \mathcal{S}}{2 \sqrt{2 \mathcal{S}}} \right) \right]^{-1}
\label{eq:norm_n}
\end{equation}
and
\begin{equation}
C = \frac{\exp[(\alpha+1) \alpha \mathcal{S} / 2]}{\sqrt{2\pi \mathcal{S}}},
\label{eq:norm_c}
\end{equation}
under the assumptions that $P_{M}(y)$ is properly normalized, continuous, and differentiable. Note that the different signs in Equations~(\ref{eq:y_trans})--(\ref{eq:norm_c}) from the original equations in \citet{Burkhart2018} are due to our adoption of the mass-weighted instead of volume-weighted PDF. 

We show in Figure~\ref{fig:pdfs_fits_more} a comparison between the best-fit log-normal gas density PDFs across the halo with and without allowing for the power-law modification specified by Equations~(\ref{eq:y_trans})--(\ref{eq:norm_c}). As suggested by the best-fit values of $\alpha$ and $\mathcal{S}$, at small radii the same $\mathcal{S}$ value is inferred in either case and the piecewise distribution tends to have a large $\alpha$, which implies a transition to a steep power-law tail at very high densities. On the contrary, at large radii the true distribution is slightly better characterized by a piecewise distribution, with an $\mathcal{S}$ different from that in the case of pure log-normal distribution and a smaller $\alpha$ corresponding to a transition to a shallower power law at low densities. Since the majority of star formation occurs in the inner halo, we focus on the simpler log-normal density PDF for the results in the main text. 

\begin{figure*}
 \centering
 \includegraphics[width=\textwidth]{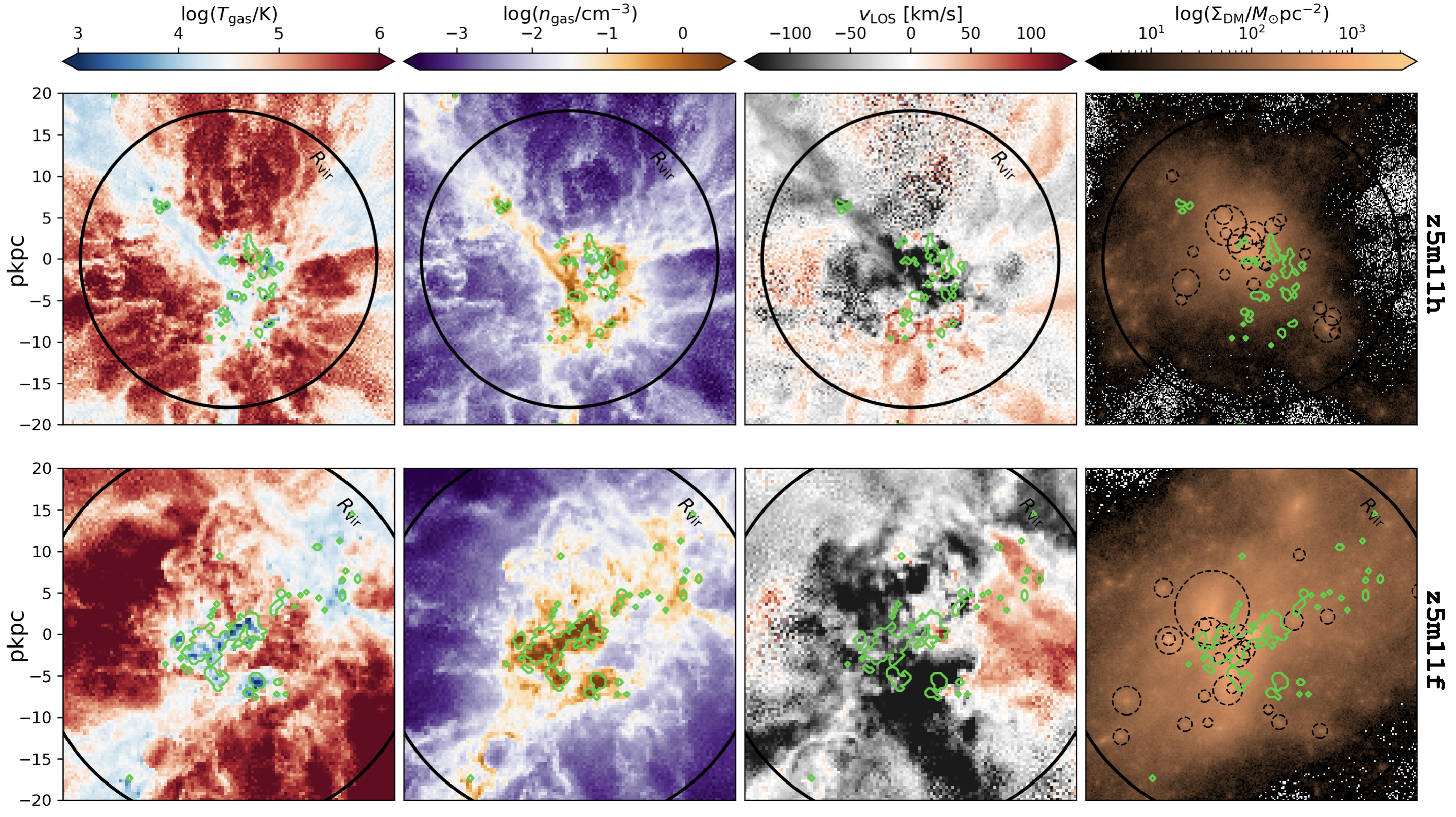}
 \caption{Gas temperature, number density, line-of-sight velocity, and dark matter density projections of the simulations $\texttt{z5m11h}$ and $\texttt{z5m11f}$ at $z=6$. Locations of subhalos with $M_\mathrm{vir}>10^8\,M_{\odot}$ are marked in the rightmost panels by dashed circles, the size of which represents $1/3R_\mathrm{vir,sub}$. Distributions of extended recent star formation (green contours, see also Figure~\ref{fig:maps}) and subhalos show no significant correlation.}
 \label{fig:subhalo_vis}
\end{figure*}

\begin{figure*}
 \centering
 \includegraphics[width=\textwidth]{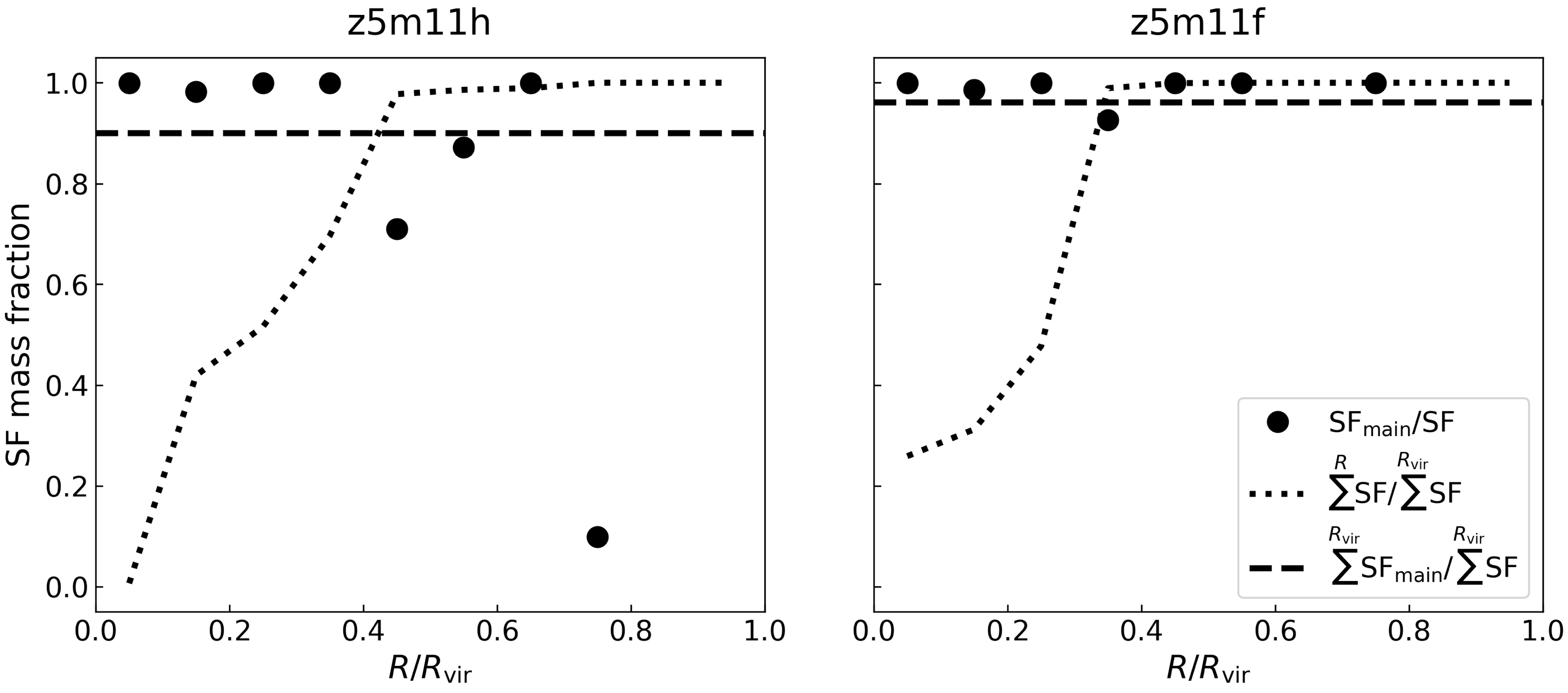}
 \caption{Comparison of recent star formation associated with subhalos in the simulations $\texttt{z5m11h}$ and $\texttt{z5m11f}$ at $z=6$. Across the galaxy's host halo, the mass fraction of recently formed, $t_\mathrm{age}<10\,$Myr star particles in only the main halo (i.e., not within 1/3$R_\mathrm{vir,sub}$ of any subhalo more massive than $3\times10^{7}\,M_{\odot}$) shown by the black circles for individual radial bins remains high at most radii, with the horizontal dashed line indicating the average of all the radial bins. The main halo has recent star formation in significant amounts out to $R\sim0.3$--0.4\,$R_\mathrm{vir}$, as illustrated by the dotted curves. }
 \label{fig:subhalo_frac}
\end{figure*}

\begin{figure*}
 \centering
 \includegraphics[width=0.95\textwidth]{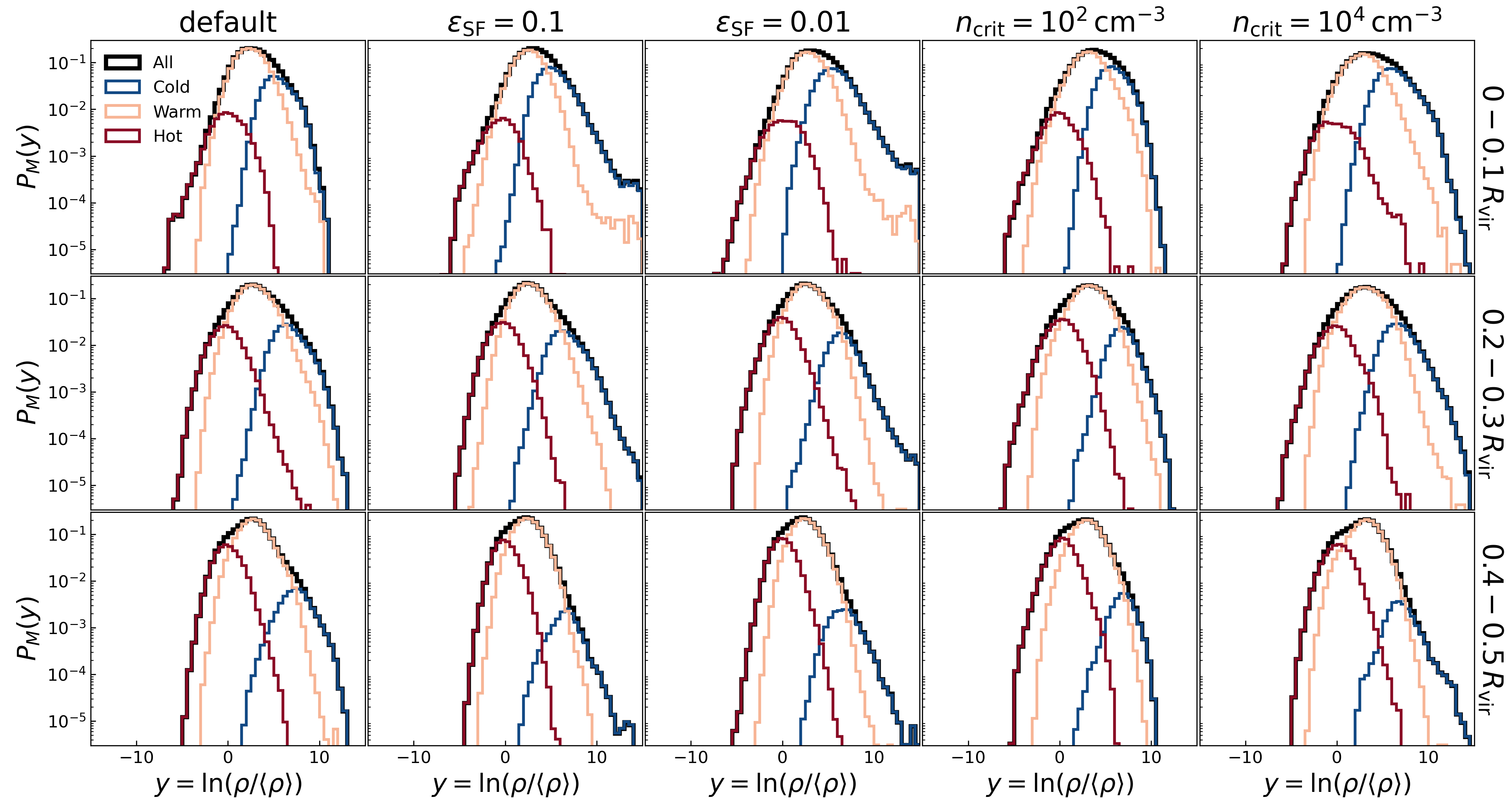}
 \caption{Density contrast PDFs of cold, warm, and hot gas at different halo radii measured for reruns of the simulation $\texttt{z5m11h}$ assuming the default FIRE-2 physics ($\epsilon_\mathrm{SF}=1$ and $n_\mathrm{crit}=10^3\,\mathrm{cm^{-3}}$), or when either the per-free-fall-time SFE or the critical density for star formation is set to a different value. While basic characteristics of the total PDF, including its single peak and the dichotomy between non-star-forming and star-forming regimes, remain unchanged, noticeable differences show up at the dense end.}
 \label{fig:model_var_phase}
\end{figure*}

\begin{figure*}
 \centering
 \includegraphics[width=0.95\textwidth]{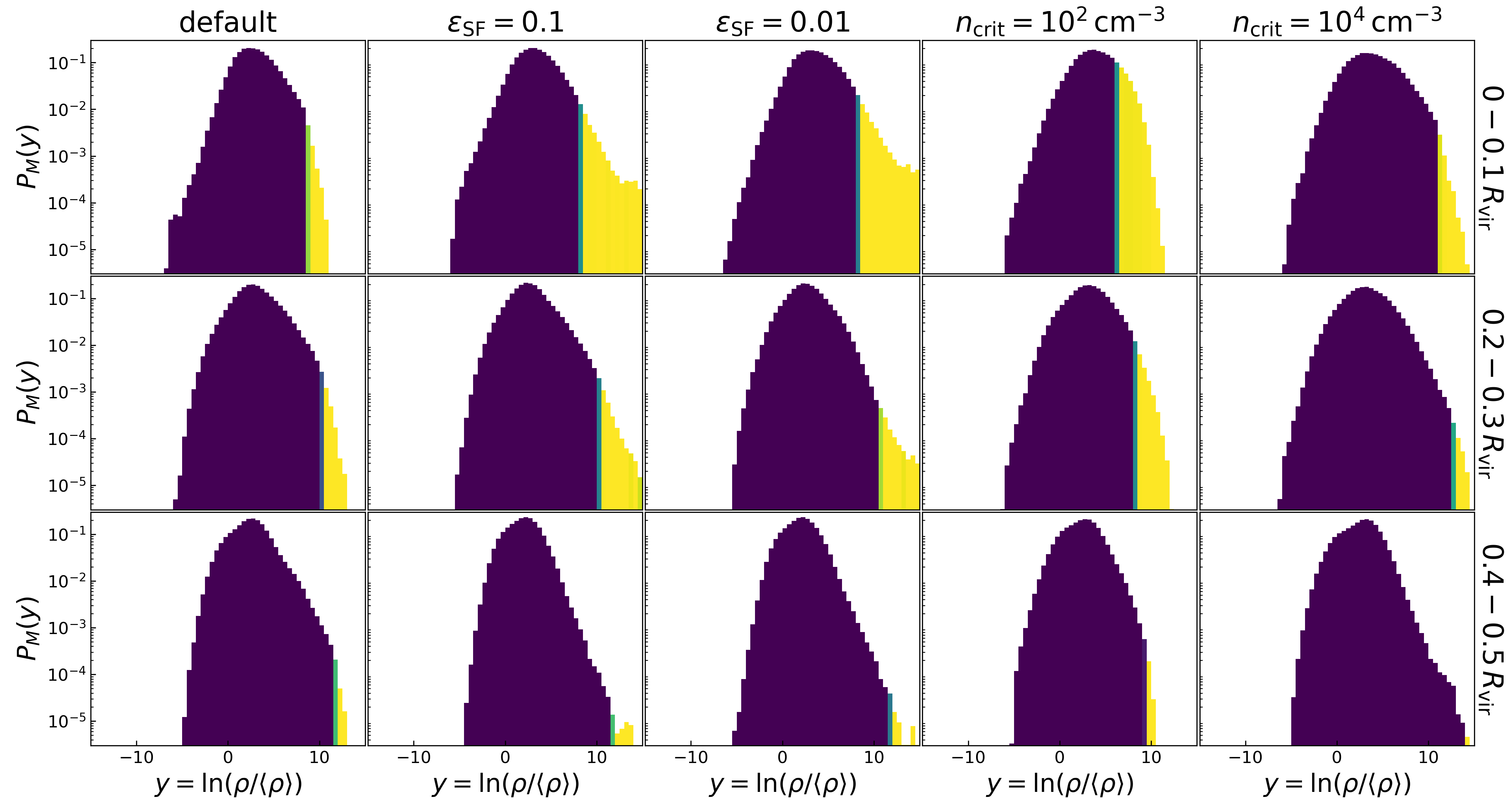}
 \caption{The same as Figure~\ref{fig:model_var_phase}, but with the total PDF color-coded by the mass fraction of star-forming gas (gas particles with $\mathrm{SFR}>0$) using the same color scheme as in Figure~\ref{fig:pdfs_fits_more}. Artificially lowering the efficiency $\epsilon_\mathrm{SF}$ populates and further extends the densest end of the PDF by causing more dense, star-forming gas to pile up, whereas lifting (reducing) the density threshold $n_\mathrm{crit}$ increases (decreases) the overall width of the PDF.}
 \label{fig:model_var_sf_frac}
\end{figure*}

\section{Star Formation in Subhalos} \label{sec:appendix-subhalo}

A valid question regarding the spatially extended star formation seen in our FIRE simulations is whether, and to what extent, it may be associated with the contribution from subhalos. To test this, in Figure~\ref{fig:subhalo_vis} we show for the simulations \texttt{z5m11h} and \texttt{z5m11f} the locations of subhalos with $M_\mathrm{vir}$ greater than $10^8\,M_{\odot}$ (corresponding to a virial temperature of $T_\mathrm{vir} \sim 10^4\,$K, below which the atomic cooling of gas becomes inefficient for stars to form) on top of the projected dark matter density, along with the same kind of gas temperature, density, and velocity distributions as displayed in Figure~\ref{fig:maps}. From these comparisons, there is no apparent correlation between where recent star formation (star particles formed in the past 10\,Myr) occurs and where subhalos capable of forming stars reside. 

We assess the subhalo contribution to spatially extended star formation at different halo radii more quantitatively in Figure~\ref{fig:subhalo_frac}, where we show the mass fraction of recent star formation associated with only the main galaxy host halo. As indicated by the black circles, only a modest fraction (less than 10\% by mass) of recent star formation, which occurs in significant amounts out to approximately $0.3$--0.4\,$R_\mathrm{vir}$ (dotted curve), can be attributed to subhalos.  

\section{Density PDFs in FIRE Model Variants} \label{sec:appendix-altmodels}

In the main text, we compare our turbulent framework for high-$z$ SFGs against the FIRE simulations created with the standard FIRE-2 physics. Here we test how the key aspects of this turbulent framework might depend on some of the important subgrid assumptions made in these simulations. In particular, modeling star formation at the resolution scale involves explicitly assumed density threshold ($n_\mathrm{crit}$) and efficiency ($\epsilon_\mathrm{SF}$) values in the FIRE-2 model. While previous studies of turbulent galaxy disks have argued that these small-scale star formation model prescriptions do not affect the predicted galaxy-scale SFRs \citep[e.g.,][]{CAFG2013} due to self-regulation by feedback, it is known that exact choices of $n_\mathrm{crit}$ and $\epsilon_\mathrm{SF}$ do impact the predicted properties of dense gas \citep{ShettyOstriker2012,Hopkins2013,Hopkins2018}. A thorough study of whether high-$z$, turbulence-dominated SFGs are consistent with these findings is beyond the scope of this work. To build up some intuition for the consequences of star formation model choices, we compare the density contrast PDF in a few variants of the standard FIRE-2 star formation model, where either $n_\mathrm{crit}$ or $\epsilon_\mathrm{SF}$ is adjusted from the default value. We ran these model variants for the simulation $\texttt{z5m11h}$, keeping everything else fixed. 

In Figures~\ref{fig:model_var_phase} and \ref{fig:model_var_sf_frac}, we present characterizations of the density contrast PDF similar to Figures~\ref{fig:pdfs_phases} and \ref{fig:pdfs_fits}, but here for the different model variants. These comparisons reveal that key qualitative features of the total PDF, such as its broad single peak and the clear dichotomy between non-star-forming and star-forming regimes, are robust across a range of model assumptions. At the same time, the comparisons highlight noticeable differences at the dense end of the PDF that reflect the influence of the resolution-scale star formation parameters. For instance, increasing $n_\mathrm{crit}$ from $10^3\,\mathrm{cm^{-3}}$ to $10^4\,\mathrm{cm^{-3}}$ broadens the distribution of both cold and warm gas at the dense end. A similar broadening, along with a pile-up of very dense gas at the distribution tail, can be seen when $\epsilon_\mathrm{SF}$ is decreased by a factor of 10 or 100 as well. Nevertheless, star formation remains confined to the densest tail of the PDF that comprises a small fraction of the total gas mass. We have also verified that the SFH of the simulated galaxy in these model variants closely resemble that in the default case. The main exception is the case where $n_\mathrm{crit}$ is reduced to $100\,\mathrm{cm^{-3}}$, which slightly narrows the width of the PDF. In this case, the overall galaxy-scale SFR is overpredicted by more than a factor of two, suggesting that the threshold density is too low, allowing spurious star formation to occur in gas clouds that are not sufficiently dense. This discrepancy is not unexpected, however, given that a threshold of $100\,\mathrm{cm^{-3}}$ lies close to the galaxy’s mean gas density, as indicated by the color-coding in Figure~\ref{fig:model_var_sf_frac}, and is therefore not sufficiently high to model star formation at high redshift. These comparisons confirm that, despite introducing some subtle differences to the PDF, adopting different resolution-scale star formation model assumptions has a relatively modest impact on our main results, provided that physically reasonable choices are made.

\bibliography{turb}{}
\bibliographystyle{aasjournal}



\end{document}